\documentclass{ectj}

\usepackage{amsfonts,amssymb,graphics,epsfig,verbatim,bm,latexsym,amsmath,url,amsbsy,array,booktabs}

\newtheorem{proposition}{Proposition}
\newtheorem{corollary}{Corollary}

\newtheorem{algorithm}{Algorithm}
\newtheorem{definition}{Definition}

\renewcommand{\thesection}{\arabic{section}}
\renewcommand{\theequation}{\arabic{section}.\arabic{equation}}

\newcounter{bean}

\received{September 2021}
\accepted{...}
\volume{21}

\title[Algorithms for Inference in SVARs]{Algorithms for Inference in SVARs Identified with Sign and Zero Restrictions}

\author[M. Read]{Matthew Read$^{\dagger}$}

\address{$^{\dagger}$Department of Economics, University College London, 30 Gordon St, London, \\
WC1H 0AX, United Kingdom \\
Reserve Bank of Australia, 65 Martin Place, Sydney, New South Wales, 2000, Australia.}
\email{matthew.read.16@ucl.ac.uk}

\def\AmSTeX{$\cal A$\kern-.1667em\lower.5ex\hbox{$\cal M$}\kern-.125em
    $\cal S$-\TeX}
\def\BibTeX{{\rm B\kern-.05em{\sc i\kern-.025em b}\kern-.08em
    T\kern-.1667em\lower.7ex\hbox{E}\kern-.125emX}}

\begin{document}
\bibliographystyle{chicago}

    \begin{abstract}
        I develop algorithms to facilitate Bayesian inference in structural vector autoregressions that are set-identified with sign and zero restrictions by showing that the system of restrictions is equivalent to a system of sign restrictions in a lower-dimensional space. Consequently, algorithms applicable under sign restrictions can be extended to allow for zero restrictions. Specifically, I extend algorithms proposed in \cite{Amir-Ahmadi_Drautzburg_2021} to check whether the identified set is nonempty and to sample from the identified set without rejection sampling. I compare the new algorithms to alternatives by applying them to variations of the model considered by \citet{Arias_Caldara_Rubio-Ramirez_2019}, who estimate the effects of US monetary policy using sign and zero restrictions on the monetary policy reaction function. The new algorithms are particularly useful when a rich set of sign restrictions substantially truncates the identified set given the zero restrictions.

        \keywords{Bayesian inference, Set identification, Sign and zero restrictions, Structural vector autoregression.}

    \end{abstract}


    \section{Introduction}

    \label{sec:introduction}

\setcounter{equation}{0}

Structural vector autoregressions (SVARs) are used in macroeconomics to estimate the dynamic causal effects of structural shocks. Parameters in these models have traditionally been point-identified using zero restrictions on the SVAR's structural parameters. However, it has become increasingly common to set-identify parameters using sign restrictions and/or a set of zero restrictions that are insufficient to achieve point-identification.\footnote{For an overview of identification in SVARs, see \cite{Kilian_Lutkepohl_2017} or \cite{Stock_Watson_2016}.} Inference in set-identified SVARs has typically been carried out via Bayesian methods that rely on rejection sampling.\footnote{For examples of frequentist approaches to inference in set-identified SVARs, see \citet{Gafarov_Meier_Montiel-Olea_2018} and \citet{Granziera_Moon_Schorfheide_2018}.} For example, when there are both sign and zero restrictions, the algorithms in Arias et al. (2018) involve drawing parameter values satisfying the zero restrictions and discarding them if they do not satisfy the sign restrictions. A drawback of this approach is that it may be computationally demanding when the sign restrictions considerably truncate the identified set given the zero restrictions, because many draws of the parameters satisfying the zero restrictions may be required to obtain a sufficient number of draws that additionally satisfy the sign restrictions. To address this problem, this paper develops algorithms to facilitate Bayesian inference in SVARs that are set-identified using a combination of sign and zero restrictions.

It is convenient to parameterize set-identified SVARs in terms of the VAR's reduced-form parameters and an orthonormal matrix so that sign and zero restrictions can be expressed as restrictions on this matrix. I focus on the case where sign and zero restrictions linearly constrain a single column of the orthonormal matrix, which I denote by $\mathbf{q} \in \mathbb{R}^{n}$, where $n$ is the dimension of the VAR. I also discuss extensions to the case where there are restrictions on multiple columns of the orthonormal matrix. The algorithms I develop are compatible with a wide range of sign and zero restrictions. These restrictions include sign restrictions on impulse responses (e.g. \citealp{Uhlig_2005}), bounds on elasticities (e.g. \citealp{Kilian_Murphy_2012}) and shape restrictions (e.g. \citealp{Amir-Ahmadi_Drautzburg_2021}), as well as sign and zero restrictions on the structural parameters themselves. Other compatible zero restrictions include `short-run' restrictions on impact impulse responses, as in \citet{Christiano_Eichenbaum_Evans_1999} or \cite{Sims_1980}, `long-run' restrictions, as in \cite{Blanchard_Quah_1989}, and restrictions arising from external instruments or `proxies', as in \cite{Mertens_Ravn_2013} and \cite{Stock_Watson_2018}. The algorithms can also accommodate certain types of `narrative restrictions', including restrictions on the sign of a structural shock in particular periods (e.g. \citealp{Antolin-Diaz_Rubio-Ramirez_2018}) or the timing of its maximum realization (e.g. \citealp{Giacomini_Kitagawa_Read_2021b}).

The algorithms developed in this paper build on those proposed in \cite{Amir-Ahmadi_Drautzburg_2021} (AD21), which are applicable when there are sign restrictions only. AD21 show that the problem of determining whether the identified set is nonempty can be cast as a linear program, which can be solved rapidly using standard software. They also propose a Gibbs sampler that draws $\mathbf{q}$ from a uniform distribution conditional on sign restrictions. Importantly, both algorithms avoid rejection sampling and so may be more computationally efficient than existing algorithms when the sign restrictions substantially truncate the identified set given the zero restrictions.\footnote{\citet{Rubio-Ramirez_Waggoner_Zha_2010} describe an algorithm for drawing from a uniform distribution over the space of orthonormal matrices conditional on sign restrictions. \cite{Arias_Rubio-Ramirez_Waggoner_2018} extend this algorithm to allow for zero restrictions. Both algorithms use rejection sampling to impose that the draws satisfy the sign restrictions.}

To extend the algorithms in AD21 to allow for zero restrictions, I show how a system of sign and zero restrictions in $\mathbb{R}^{n}$ can be expressed as an equivalent system of sign restrictions in a lower-dimensional space. The algorithms in AD21 are applicable to the transformed system of sign restrictions and, in conjunction with a simple transformation, can be used to obtain values of the parameters satisfying the original identifying restrictions. Specifically, an algorithm determines whether the identified set is nonempty by solving a linear program and, if so, generates a value of $\mathbf{q}$ satisfying the identifying restrictions. This value of $\mathbf{q}$ can be used to initialize a Gibbs sampler that draws from a uniform distribution over the identified set for $\mathbf{q}$. Additionally, it can be used to initialize a gradient-based numerical optimization routine whose aim is to compute the bounds of the identified set for a scalar parameter of interest, which is useful in the context of prior-robust Bayesian inference (e.g. \citealp{Giacomini_Kitagawa_2021}).

I illustrate the algorithms using the empirical application in \citet{Arias_Caldara_Rubio-Ramirez_2019} (ACR19). They estimate the effects of monetary policy shocks in the United States by imposing sign and zero restrictions on the monetary policy reaction function. I augment these restrictions with the sign restrictions on impulse responses considered in \cite{Uhlig_2005}, and explore the accuracy and computational efficiency of my algorithms relative to alternatives. My algorithms are particularly useful when a large number of sign restrictions appreciably truncate the identified set given the zero restrictions. These algorithms should therefore facilitate the use of rich sets of sign restrictions alongside zero restrictions.

As an additional illustration of the utility of the algorithms, I impose a restriction on the timing of the maximum realization of the monetary policy shock considered in \citet{Giacomini_Kitagawa_Read_2021b}; in addition to the restrictions from ACR19 and \citet{Uhlig_2005}, I impose that the monetary policy shock in October 1979 -- the month in which Paul Volcker dramatically and unexpectedly raised the federal funds rate -- was the largest positive realization of the shock in the sample. This restriction generates as many sign restrictions as observations in the sample (around 500 in this application) and tends to truncate the identified set considerably, which makes existing algorithms extremely computationally burdensome. Under this restriction, output falls with high posterior probability following a positive monetary policy shock. However, the identified set is empty in around 95~per cent of draws from the posterior of the reduced-form parameters, which suggests the restriction is inconsistent with the data.

\medskip

\noindent\textbf{Outline.} The remainder of the paper is structured as follows. Section~\ref{sec:svar} outlines the SVAR framework and describes the identifying restrictions considered. Section~\ref{sec:algorithms} shows how a system of sign and zero restrictions can be expressed as an equivalent system of sign restrictions in a lower-dimensional space and explains how algorithms used to conduct inference in sign-restricted SVARs can consequently be extended to the case of zero restrictions. Section~\ref{sec:numerical} describes how to numerically implement the algorithms. Section~\ref{sec:empirical} explores the accuracy and efficiency of the algorithms relative to existing alternatives using the model in ACR19 augmented with additional identifying restrictions. Section~\ref{sec:extensions} discusses extending the algorithms to allow for restrictions on additional columns of the orthonormal matrix. Section~\ref{sec:conclusion} concludes. Proofs are relegated to the Appendix.

\medskip

\noindent\textbf{Generic notation.} For a matrix $\mathbf{X}$, $\mathrm{vec}(\mathbf{X})$ is the vectorization of $\mathbf{X}$. When $\mathbf{X}$ is symmetric, $\mathrm{vech}(\mathbf{X})$ is the half-vectorization of $\mathbf{X}$, which stacks the elements of $\mathbf{X}$ lying on or below the diagonal into a vector. $\mathbf{e}_{i,n}$ is the $i$th column of the $n\times n$ identity matrix, $\mathbf{I}_{n}$. $\mathbf{0}_{m\times n}$ is an $m\times n$ matrix of zeros. $\lVert . \rVert$ is the Euclidean norm. $\mathbb{S}^{n-1}$ is the unit sphere in $\mathbb{R}^{n}$ (i.e. the set $\{\mathbf{q} \in \mathbb{R}^{n}: \mathbf{q}'\mathbf{q} = 1\}$).

\section{Framework}
\label{sec:svar}

\setcounter{equation}{0}

\subsection{SVAR}

Let $\mathbf{y}_{t}$ be an $n\times 1$ vector of endogenous variables following the SVAR($p$) process:
\begin{equation}\label{eq:SVARp}
    \mathbf{A}_{0}\mathbf{y}_{t} = \sum_{l=1}^{p}\mathbf{A}_{l}\mathbf{y}_{t-l} + \bm{\varepsilon}_{t}, \quad t=1,...,T,
\end{equation}
where $\mathbf{A}_{0}$ is invertible and $\bm{\varepsilon}_{t}\overset{iid}{\sim} N(\mathbf{0}_{n\times 1},\mathbf{I}_{n})$ are structural shocks. The diagonal elements of $\mathbf{A}_{0}$ are normalized to be positive, which is a normalization on the signs of the structural shocks. Exogenous regressors (such as a constant) are omitted for simplicity of exposition, but these are straightforward to include. Letting $\mathbf{x}_{t} =
(\mathbf{y}_{t-1}',\ldots,\mathbf{y}_{t-p}')'$ and $\mathbf{A}_{+} = (\mathbf{A}_{1},\ldots,\mathbf{A}_{p})$, rewrite the SVAR($p$) as
\begin{equation}\label{eq:SVARpCondensed}
  \mathbf{A}_{0}\mathbf{y}_{t} = \mathbf{A}_{+}\mathbf{x}_{t} + \bm{\varepsilon}_{t}, \quad t=1,...,T.
\end{equation}
$(\mathbf{A}_{0},\mathbf{A}_{+})$ are the structural parameters. The reduced-form VAR($p$) representation is
\begin{equation}\label{eq:VARpCondensed}
  \mathbf{y}_{t} = \mathbf{B}\mathbf{x}_{t} + \mathbf{u}_{t}, \quad t=1,...,T,
\end{equation}
where $\mathbf{B} = (\mathbf{B}_{1},\ldots,\mathbf{B}_{p})$, $\mathbf{B}_{l}=\mathbf{A}_{0}^{-1}\mathbf{A}_{l}$ for $l=1,\ldots,p$, and $\mathbf{u}_{t} = \mathbf{A}_{0}^{-1}\bm{\varepsilon}_{t} \overset{iid}{\sim}
N(\mathbf{0}_{n\times 1},\bm{\Sigma})$ with $\bm{\Sigma} = \mathbf{A}_{0}^{-1}(\mathbf{A}_{0}^{-1})'$. $\bm{\phi} = (\mathrm{vec}(\mathbf{B})',\mathrm{vech}(\bm{\Sigma})')' \in \bm{\Phi}$ are the reduced-form parameters. The SVAR's orthogonal reduced form is
\begin{equation}\label{eq:orthogonalreducedform}
  \mathbf{y}_{t} = \mathbf{B}\mathbf{x}_{t} + \bm{\Sigma}_{tr}\mathbf{Q}\bm{\varepsilon}_{t}, \quad t=1,...,T,
\end{equation}
where $\bm{\Sigma}_{tr}$ is the lower-triangular Cholesky factor of $\bm{\Sigma}$ (i.e. $\bm{\Sigma}_{tr}\bm{\Sigma}_{tr}'=\bm{\Sigma}$) with nonnegative diagonal elements and $\mathbf{Q}$ is an $n\times n$ orthonormal matrix with $j$th column $\mathbf{q}_{j}$. Let $\mathcal{O}(n)$ denote the space of all $n\times n$ orthonormal matrices.

Impulse responses are typically the parameters of interest in analyses using SVARs. The horizon-$h$ impulse response of the $i$th variable to the $j$th shock is
\begin{equation}
    \eta_{i,j,h} \equiv \eta_{i,j,h}(\bm{\phi},\mathbf{q}_{j}) = \mathbf{e}_{i,n}'\mathbf{C}_{h}\bm{\Sigma}_{tr}\mathbf{q}_{j},
\end{equation}
where $\mathbf{C}_{h}$ is defined recursively by $\mathbf{C}_{h} = \sum_{l=1}^{\min\left\{h,p\right\}}\mathbf{B}_{l}\mathbf{C}_{h-l}$ for $h \geq 1$ with $\mathbf{C}_{0} = \mathbf{I}_{n}$.

\subsection{Identifying restrictions}

Consider the case where there are linear sign and zero restrictions constraining $\mathbf{q}_{1}$ only (extensions to this case are discussed in Section~\ref{sec:extensions}). Let $\mathbf{F}(\bm{\phi})$ be the $r\times n$ matrix whose rows represent the coefficients of $r$ zero restrictions, so $\mathbf{F}(\bm{\phi})\mathbf{q}_{1} = \mathbf{0}_{r\times 1}$. For example, zero restrictions on the first row of $\mathbf{A}_{0}$ take the form $\mathbf{e}_{1,n}'\mathbf{A}_{0}\mathbf{e}_{i,n} = (\bm{\Sigma}_{tr}^{-1}\mathbf{e}_{i,n})'\mathbf{q}_{1} = 0$, zero restrictions on impact responses to the first shock take the form $\mathbf{e}_{i,n}'\mathbf{A}_{0}^{-1}\mathbf{e}_{1,n} = \mathbf{e}_{i,n}'\bm{\Sigma}_{tr}\mathbf{q}_{1} = 0$ and long-run restrictions on cumulative impulse responses to the first shock take the form $\mathbf{e}_{i,n}'(\mathbf{I}_{n}-\sum_{l=1}^{p}\mathbf{B}_{l})^{-1}\bm{\Sigma}_{tr}\mathbf{q}_{1} = 0$. Assume $0 < r < n-1$, which implies $\mathbf{q}_{1}$ is set-identified (\citealp{Rubio-Ramirez_Waggoner_Zha_2010}), and assume $\mathrm{rank}(\mathbf{F}(\bm{\phi})) = r$.

Similarly, let $\mathbf{S}(\bm{\phi})\mathbf{q}_{1} \geq \mathbf{0}_{s\times 1}$ represent a set of $s$ sign restrictions, which includes the sign normalization $\mathbf{e}_{1,n}'\mathbf{A}_{0}\mathbf{e}_{1,n} = (\bm{\Sigma}_{tr}^{-1}\mathbf{e}_{1,n})'\mathbf{q}_{1} \geq 0$. $\mathbf{S}(\bm{\phi})$ may include restrictions on impulse responses to a standard-deviation shock, ratios of these impulse responses (e.g. elasticity and shape restrictions) and/or elements of the first row of $\mathbf{A}_{0}$. For example, a bound on the impact impulse response of the $i$th variable to a shock in the first variable that raises the first variable by one unit is $(\mathbf{e}_{i,n}'\bm{\Sigma}_{tr}\mathbf{q}_{1})/(\mathbf{e}_{1,n}'\bm{\Sigma}_{tr}\mathbf{q}_{1}) \geq \lambda$, where $\lambda$ is a known scalar. This restriction can be expressed as $(\mathbf{e}_{i,n}'-\lambda\mathbf{e}_{1,n}')\bm{\Sigma}_{tr}\mathbf{q}_{1} \geq 0$. An example of a shape restriction is that the horizon-$h$ impulse response of the first variable to the first shock is weakly greater than the horizon-$l$ response, which requires that $\mathbf{e}_{i,n}'\mathbf{C}_{h}\bm{\Sigma}_{tr}\mathbf{q}_{1} \geq \mathbf{e}_{i,n}'\mathbf{C}_{l}\bm{\Sigma}_{tr}\mathbf{q}_{1}$ or  $\mathbf{e}_{i,n}'(\mathbf{C}_{h}-\mathbf{C}_{l})\bm{\Sigma}_{tr}\mathbf{q}_{1} \geq 0$. $\mathbf{S}(\bm{\phi})$ may also include particular types of narrative restrictions, including restrictions on the sign or relative magnitude of the first shock in particular periods. For example, the restriction that the first shock is nonnegative in period $k$ is $\mathbf{e}_{1,n}'\mathbf{A}_{0}\mathbf{u}_{k} = (\bm{\Sigma}_{tr}^{-1}\mathbf{u}_{k})'\mathbf{q}_{1} \geq 0$. The restriction that the first shock in period $k$ is larger than the first shock in period $m$ is $\mathbf{e}_{1,n}'\mathbf{A}_{0}\mathbf{u}_{k} \geq \mathbf{e}_{1,n}'\mathbf{A}_{0}\mathbf{u}_{m}$, which is equivalent to $(\bm{\Sigma}_{tr}^{-1}(\mathbf{u}_{k}-\mathbf{u}_{m}))'\mathbf{q}_{1} \geq 0$.\footnote{Under narrative restrictions, $\mathbf{S}(\bm{\phi})$ is also a function of the data in particular periods through the reduced-form innovations, but I leave this potential dependence implicit. See \cite{Antolin-Diaz_Rubio-Ramirez_2018} or \cite{Giacomini_Kitagawa_Read_2021b} for further details on narrative restrictions.}

Given a set of identifying restrictions, the identified set for $\mathbf{q}_{1}$ collects observationally equivalent parameter values and is defined as
\begin{equation}
    \mathcal{Q}_{1}(\bm{\phi}|\mathbf{F},\mathbf{S}) = \left\{\mathbf{q}_{1} \in \mathbb{S}^{n-1}: \mathbf{F}(\bm{\phi})\mathbf{q}_{1} = \mathbf{0}_{r\times 1}, \mathbf{S}(\bm{\phi})\mathbf{q}_{1} \geq \mathbf{0}_{s\times 1}\right\}.
\end{equation}
This set has a geometric interpretation. The zero restrictions $\mathbf{F}(\bm{\phi})\mathbf{q}_{1} = \mathbf{0}_{r\times 1}$ restrict $\mathbf{q}_{1}$ to lie in an $(n-r)$-dimensional hyperplane, while the sign restrictions $\mathbf{S}(\bm{\phi})\mathbf{q}_{1} \geq \mathbf{0}_{s\times 1}$ restrict $\mathbf{q}_{1}$ to lie within the intersection of $s$ half-spaces. Since $\mathbf{q}_{1}$ is a column of an orthonormal matrix, it must have unit length, so it lies on the unit sphere in $\mathbb{R}^{n}$. The identified set for $\mathbf{q}_{1}$ is the intersection of these spaces, which may be empty at particular values of $\bm{\phi}$. The identified set for an impulse response $\eta_{i,j,h}$ is
\begin{equation}
  IS_{\eta}(\bm{\phi}|\mathbf{F},\mathbf{S}) = \left\{\eta_{i,j,h}(\bm{\phi},\mathbf{q}_{1}): \mathbf{q}_{1} \in \mathcal{Q}_{1}(\bm{\phi}|\mathbf{F},\mathbf{S})\right\}.
\end{equation}

\section{Transforming the System of Identifying Restrictions}
\label{sec:algorithms}

\setcounter{equation}{0}

This section shows that the system of equality and inequality restrictions in $\mathbb{R}^{n}$ can be expressed as an equivalent system of inequality restrictions in $\mathbb{R}^{n-r}$. Subsequently, I explain how the algorithms proposed in AD21 for the case of sign restrictions can be used to check whether $\mathcal{Q}_{1}(\bm{\phi}|\mathbf{F},\mathbf{S})$ is nonempty and, if so, to generate a value of $\mathbf{q}_{1}$ satisfying the identifying restrictions. Additionally, the Gibbs sampler developed in AD21 can be extended to randomly sample $\mathbf{q}_{1}$ from a uniform distribution over $\mathcal{Q}_{1}(\bm{\phi}|\mathbf{F},\mathbf{S})$.

Let $N(\mathbf{F}(\bm{\phi}))$ denote an orthonormal basis for the null space of $\mathbf{F}(\bm{\phi})$, which spans the hyperplane $\mathbf{F}(\bm{\phi})\mathbf{q}_{1} = \mathbf{0}_{r\times 1}$. Under the assumption $\mathrm{rank}(\mathbf{F}(\bm{\phi})) = r$, the rank-nullity theorem implies $N(\mathbf{F}(\bm{\phi}))$ is of dimension $n-r$. The null space of $N(\mathbf{F}(\bm{\phi}))'$ is then of dimension $r$ and the columns of the matrix $\mathbf{K} = (N(\mathbf{F}(\bm{\phi})),N(N(\mathbf{F}(\bm{\phi}))'))$ form an orthonormal basis for $\mathbb{R}^{n}$.\footnote{I leave the dependence of $\mathbf{K}$ on $\mathbf{F}(\bm{\phi})$ implicit.} The matrix that transforms from this basis into the standard basis is $\mathbf{K}^{-1}$. In the new basis, the coefficients in the zero and sign restrictions are, respectively, $\tilde{\mathbf{F}}(\bm{\phi}) = (\mathbf{K}^{-1}\mathbf{F}(\bm{\phi})')'$ and $\tilde{\mathbf{S}}(\bm{\phi}) = (\mathbf{K}^{-1}\mathbf{S}(\bm{\phi})')'$. After applying this change of basis, the hyperplane generated by the zero restrictions coincides with the hyperplane spanned by the first $n-r$ basis vectors (i.e. the first $n-r$ column vectors of $\mathbf{I}_{n}$). Any vector lying in this hyperplane will therefore have its last $r$ elements equal to zero.

After the change of basis, the projection of the $i$th row of $\tilde{\mathbf{S}}(\bm{\phi})$, $\tilde{\mathbf{S}}_{i}(\bm{\phi})$, onto the hyperplane generated by the zero restrictions is\footnote{Since the hyperplane generated by the zero restrictions coincides with the hyperplane spanned by the first $n-r$ basis vectors, this is equivalent to projecting onto the linear subspace spanned by the first $n-r$ basis vectors via $\bar{\mathbf{S}}_{i}(\bm{\phi})' = (\mathbf{I}_{n} - \mathbf{B}(\mathbf{B}'\mathbf{B})^{-1}\mathbf{B}')\tilde{\mathbf{S}}_{i}(\bm{\phi})'$, where $\mathbf{B} = (\mathbf{0}_{r\times (n-r)},\mathbf{I}_{r})'$ contains the last $r$ basis vectors.}
\begin{equation}\label{eq:projection}
  \bar{\mathbf{S}}_{i}(\bm{\phi})' = (\mathbf{I}_{n} - \tilde{\mathbf{F}}(\bm{\phi})'(\tilde{\mathbf{F}}(\bm{\phi})\tilde{\mathbf{F}}(\bm{\phi})')^{-1}\tilde{\mathbf{F}}(\bm{\phi}))\tilde{\mathbf{S}}_{i}(\bm{\phi})'.
\end{equation}
Let $\mathbf{M} = (\mathbf{I}_{n-r},\mathbf{0}_{(n-r)\times r})$ be the $(n-r)\times n$ matrix such that $\mathbf{M}\mathbf{x}$ drops the last $r$ elements of the $n \times 1$ vector $\mathbf{x}$ and let $\bar{\mathbf{S}}(\bm{\phi}) = (\mathbf{M}\bar{\mathbf{S}}_{i}(\bm{\phi})',\ldots,\mathbf{M}\bar{\mathbf{S}}_{s}(\bm{\phi})')'$. The end result of these transformations is that the sign and zero restrictions in $\mathbb{R}^{n}$ have been replaced with an equivalent system of sign restrictions $\bar{\mathbf{S}}(\bm{\phi})\bar{\mathbf{q}}_{1} \geq \mathbf{0}_{s\times 1}$ in $\mathbb{R}^{n-r}$. This claim is formalized in the following proposition.

\begin{proposition} \label{prop:equivalence}
    Let $\mathbf{F}(\bm{\phi})\mathbf{q}_{1}  = \mathbf{0}_{r\times 1}$ be a system of $r$ zero restrictions with $\mathrm{rank}(\mathbf{F}(\bm{\phi})) = r$ and let $\mathbf{S}(\bm{\phi})\mathbf{q}_{1} \geq \mathbf{0}_{s\times 1}$ be a system of $s$ sign restrictions. \\
    (a) If $\mathbf{q}_{1} \in \mathbb{R}^{n}$ satisfies $\mathbf{F}(\bm{\phi})\mathbf{q}_{1} = \mathbf{0}_{r\times 1}$ and $\mathbf{S}(\bm{\phi})\mathbf{q}_{1} \geq \mathbf{0}_{s\times 1}$, then $\bar{\mathbf{q}}_{1} = \mathbf{M}\mathbf{K}^{-1}\mathbf{q}_{1} \in \mathbb{R}^{n-r}$ satisfies $\bar{\mathbf{S}}(\bm{\phi})\bar{\mathbf{q}}_{1} \geq \mathbf{0}_{s\times 1}$. \\
    (b) If $\bar{\mathbf{q}}_{1} \in \mathbb{R}^{n-r}$ satisfies $\bar{\mathbf{S}}(\bm{\phi})\bar{\mathbf{q}}_{1} \geq \mathbf{0}_{s\times 1}$, then $\mathbf{q}_{1} = \mathbf{K}\mathbf{M}'\bar{\mathbf{q}}_{1} \in \mathbb{R}^{n}$ satisfies $\mathbf{F}(\bm{\phi})\mathbf{q}_{1} = \mathbf{0}_{r\times 1}$ and $\mathbf{S}(\bm{\phi})\mathbf{q}_{1} \geq \mathbf{0}_{s\times 1}$.
\end{proposition}

This proposition implies the following corollary relating (non)emptiness of the set $\bar{\mathcal{Q}}_{1}(\bm{\phi}|\bar{\mathbf{S}}) = \left\{\bar{\mathbf{q}}_{1} \in \mathbb{S}^{n-r-1}: \bar{\mathbf{S}}(\bm{\phi})\bar{\mathbf{q}}_{1} \geq \mathbf{0}_{s\times 1}\right\}$ to (non)emptiness of the set $\mathcal{Q}_{1}(\bm{\phi}|\mathbf{F},\mathbf{S})$.

\begin{corollary}\label{cor:nonempty}
    $\mathcal{Q}_{1}(\bm{\phi}|\mathbf{F},\mathbf{S})$ is nonempty if and only if $\bar{\mathcal{Q}}_{1}(\bm{\phi}|\bar{\mathbf{S}})$ is nonempty.
\end{corollary}

Based on the results in AD21, $\bar{\mathcal{Q}}_{1}(\bm{\phi}|\bar{\mathbf{S}})$ is nonempty if the largest ball that can be inscribed within the intersection of the $s$ half-spaces generated by the inequality restrictions $\bar{\mathbf{S}}(\bm{\phi})\bar{\mathbf{q}}_{1} \geq \mathbf{0}_{s\times 1}$ and the unit $(n-r)$-cube has positive radius. The problem of finding the radius and `Chebyshev' centre of this ball can be formulated as a linear program, which can be solved efficiently (e.g. \citealp{Boyd_Vandenberghe_2004}). If the ball has positive radius with centre $\mathbf{c} \in \mathbb{R}^{n-r}$, then $\bar{\mathbf{q}}_{1}^{(0)} = \mathbf{c}/\lVert \mathbf{c} \rVert$ satisfies $\bar{\mathbf{S}}(\bm{\phi})\bar{\mathbf{q}}_{1}^{(0)} \geq \mathbf{0}_{s\times 1}$ and lies in $\mathbb{S}^{n-r-1}$. By Proposition~\ref{prop:equivalence}(ii), $\mathbf{q}_{1}^{(0)} = \mathbf{K}\mathbf{M}'\bar{\mathbf{q}}_{1}^{(0)}$ satisfies the original set of identifying restrictions and lies in $\mathbb{S}^{n-1}$.

The Gibbs sampler in AD21 can be used to obtain a sequence of draws of $\bar{\mathbf{q}}_{1}$ from a uniform distribution over $\bar{\mathcal{Q}}_{1}(\bm{\phi}|\bar{\mathbf{S}})$ using $\bar{\mathbf{q}}_{1}^{(0)}$ to initialize the sampler.\footnote{The Gibbs sampler in AD21 builds on a Gibbs sampler developed by \cite{Li_Ghosh_2015} for sampling from a multivariate normal distribution truncated by linear inequality restrictions.} Let $\bar{\mathbf{q}}_{1}^{(k)}$ represent the $k$th draw. If $\bar{\mathbf{q}}_{1}^{(k)}$ is uniformly distributed over $\bar{\mathcal{Q}}_{1}(\bm{\phi}|\bar{\mathbf{S}})$, then $\mathbf{M}'\bar{\mathbf{q}}_{1}^{(k)}$ is uniformly distributed over $\left\{\mathbf{M}'\bar{\mathbf{q}}_{1} \in \mathbb{S}^{n-1}: \tilde{\mathbf{F}}(\bm{\phi})\mathbf{M}'\bar{\mathbf{q}}_{1} = \mathbf{0}_{r\times 1}, (\mathbf{M}'\bar{\mathbf{S}}(\bm{\phi})')'\mathbf{M}'\bar{\mathbf{q}}_{1} \geq \mathbf{0}_{s\times 1} \right\}$. Since $\mathbf{K}$ is an orthonormal matrix, $\mathbf{q}_{1}^{(k)} = \mathbf{K}\mathbf{M}'\bar{\mathbf{q}}_{1}^{(k)}$ is also uniformly distributed. Applying this transformation to each draw $\bar{\mathbf{q}}_{1}^{(k)}$ therefore yields draws $\mathbf{q}_{1}^{(k)}$ that are uniformly distributed over $\mathcal{Q}_{1}(\bm{\phi}|\mathbf{F},\mathbf{S})$. These transformed draws can be used when conducting Bayesian inference under a conditionally uniform prior for $\mathbf{q}_{1}$ given $\bm{\phi}$.\footnote{The accept-reject sampler proposed in \cite{Arias_Rubio-Ramirez_Waggoner_2018} involves rejecting joint draws of $(\bm{\phi},\mathbf{Q})$ that violate the sign restrictions. In contrast, imposing a conditionally uniform prior requires obtaining a single draw of $\mathbf{Q}$ (or $\mathbf{q}_{1}$) that satisfies the identifying restrictions at each draw of the reduced-form parameters such that the identified set is nonempty. See \cite{Uhlig_2017} for a discussion of this point.}

To provide some geometric intuition, it is useful to consider the case where $n=3$ and there is one zero restriction. The set $\bar{\mathcal{Q}}_{1}(\bm{\phi}|\bar{\mathbf{S}})$ (when it is nonempty) is an arc of the unit circle in $\mathbb{R}^{2}$. The Gibbs sampler from AD21 generates draws from a uniform distribution over this arc. Applying the transformation $\mathbf{M}'\bar{\mathbf{q}}_{1}^{(k)}$ to the draws embeds the draws on this arc as draws on an arc of the unit sphere in $\mathbb{R}^{3}$, where the arc lies within the plane perpendicular to the $z$ axis. Since $\mathbf{K}$ is orthonormal, left-multiplication by $\mathbf{K}$ rotates the draws about a particular axis of rotation. The rotation preserves the distribution of the draws on the arc, which now lies within the plane perpendicular to the vector $\mathbf{F}(\bm{\phi})'$.

\section{Numerical Implementation}
\label{sec:numerical}

\setcounter{equation}{0}

This section describes numerical algorithms to facilitate inference in SVARs identified using sign and zero restrictions. Algorithm~\ref{alg:checkis} determines whether the identified set is nonempty and, if so, generates a value of $\mathbf{q}_{1}$ satisfying the identifying restrictions. Algorithm~\ref{alg:gibbs} generates draws of $\mathbf{q}_{1}$ that are uniformly distributed over $\mathcal{Q}_{1}(\bm{\phi}|\mathbf{F},\mathbf{S})$ via Gibbs sampling. The algorithms operate given a value of $\bm{\phi}$ and can be embedded within a posterior sampler for these parameters, in which case the assumption $\mathrm{rank}(\mathbf{F}(\bm{\phi})) = r$ needs to hold $\bm{\phi}$-almost surely. For convenience, I suppress dependence on $\bm{\phi}$ in the descriptions of the algorithms below.

\begin{algorithm} \label{alg:checkis}
\textnormal{\textbf{Determining whether $\mathcal{Q}_{1}(\bm{\phi}|\mathbf{F},\mathbf{S})$ is empty.} Let $\mathbf{F}\mathbf{q}_{1} \geq \mathbf{0}_{r\times 1}$ be the set of zero restrictions and let $\mathbf{S}\mathbf{q}_{1} \geq \mathbf{0}_{s\times 1}$ be the set of sign restrictions (including the sign normalization) given $\bm{\phi}$.}
\setcounter{bean}{0}
\begin{center}
\begin{list}
{\textsc{Step} \arabic{bean}.}{\usecounter{bean}}
\item \textnormal{Compute the change-of-basis matrix $\mathbf{K} = (N(\mathbf{F}),N(N(\mathbf{F})'))$ and transform the coefficient vectors of the sign and zero restrictions into the new basis via $\tilde{\mathbf{S}} = (\mathbf{K}^{-1}\mathbf{S}')'$ and $\tilde{\mathbf{F}} = (\mathbf{K}^{-1}\mathbf{F}')'$.\footnote{In the MATLAB code accompanying the paper, I implement this step using MATLAB's `null' function, which uses the singular value decomposition to compute an orthonormal basis for the null space.}}
\item \textnormal{Project the coefficient vectors of the sign restrictions in the new basis onto the linear subspace spanned by the rows of $\tilde{\mathbf{F}}$ and drop the last $r$ elements of the resulting vectors. The transformed matrix of coefficients is $\bar{\mathbf{S}} = (\mathbf{M}(\mathbf{I}_{n} - \tilde{\mathbf{F}}'(\tilde{\mathbf{F}}\tilde{\mathbf{F}}')^{-1}\tilde{\mathbf{F}})\tilde{\mathbf{S}}')'$, where  $\mathbf{M} = (\mathbf{I}_{n-r},\mathbf{0}_{(n-r)\times r})$.}
\item \textnormal{Solve for the Chebyshev centre $\mathbf{c} = (c_{1},\ldots,c_{n-r})'$ and radius $R$ of the set $\{\bar{\mathbf{q}}_{1} \in \mathbb{R}^{n-r}: \bar{\mathbf{S}}\bar{\mathbf{q}}_{1} \geq \mathbf{0}_{s\times 1}, |\bar{q}_{1,i}| \leq 1, i=1,\ldots,n-r \}$, where $\bar{q}_{1,i}$ is the $i$th element of $\bar{\mathbf{q}}_{1}$, by solving the linear program:}
    \begin{equation*}
      \max_{\{R\geq 0, \mathbf{c}\}} R
    \end{equation*}
    \textnormal{subject to}
    \begin{align*}
      \mathbf{e}_{k,s}'\bar{\mathbf{S}}\mathbf{c} + R\lVert  \mathbf{e}_{k,s}'\bar{\mathbf{S}} \rVert &\geq 0, \quad k = 1,\ldots,s, \\
      c_{i} + R &\leq 1, \quad i = 1,\ldots,n-r, \\
      c_{i} - R &\geq -1, \quad i = 1,\ldots,n-r.
    \end{align*}
\item \textnormal{If $R > 0$, conclude $\mathcal{Q}_{1}(\bm{\phi}|\mathbf{F},\mathbf{S})$ is nonempty and compute $\bar{\mathbf{q}}_{1}^{(0)} = \mathbf{c}/\lVert \mathbf{c} \rVert$. Otherwise, conclude $\mathcal{Q}_{1}(\bm{\phi}|\mathbf{F},\mathbf{S})$ is empty.}
\end{list}
\end{center}
\end{algorithm}

If interest is in computing the bounds of the identified set for a scalar function of $\bm{\phi}$ and $\mathbf{q}_{1}$, such as $\eta_{i,j,h}$, $\mathbf{q}_{1}^{(0)} = \mathbf{K}\mathbf{M}'\bar{\mathbf{q}}_{1}^{(0)}$ is a feasible value of $\mathbf{q}_{1}$ satisfying the identifying restrictions and can be used to initialize a gradient-based optimization algorithm. This is relevant in the context of conducting prior-robust Bayesian inference, as in \cite{Giacomini_Kitagawa_2021}.

If interest is in obtaining uniformly distributed draws over $\mathcal{Q}_{1}(\bm{\phi}|\mathbf{F},\mathbf{S})$, $\bar{\mathbf{q}}_{1}^{(0)}$ can be used to initialize the following Gibbs sampler.

\begin{algorithm} \label{alg:gibbs}
\textnormal{\textbf{Gibbs sampler for $\mathbf{q}_{1}$.} Assume the output of Algorithm~\ref{alg:checkis} is available and $\mathcal{Q}_{1}(\bm{\phi}|\mathbf{F},\mathbf{S})$ is nonempty. Initialize the algorithm at $\mathbf{z}^{(0)} = \bar{\mathbf{q}}_{1}^{(0)}$ and let $L$ be the desired number of draws of $\mathbf{q}_{1}$. For $k=1,\ldots,L$, iterate on the following steps:}
\setcounter{bean}{0}
\begin{center}
\begin{list}
{\textsc{Step} \arabic{bean}.}{\usecounter{bean}}
\item \textnormal{Let $\bar{\mathbf{S}}_{j,v:w}$ be elements $v,v+1,\ldots,w-1,w$ of the $j$th row of $\bar{\mathbf{S}}$ and let $\mathbf{z}_{v:w}^{(k)}$ be elements $v,v+1,\ldots,w-1,w$ of $\mathbf{z}^{(k)}$. For $i = 1,\ldots,n-r$, draw $z_{i}^{(k)}$ from the truncated standard normal distribution with lower bound $l_{i}^{(k)}$ and upper bound $u_{i}^{(k)}$, where}
      \begin{align*}
        l_{i}^{(k)} &= \max\Big\{-\frac{\bar{\mathbf{S}}_{j,1:(i-1)}\mathbf{z}_{1:(i-1)}^{(k)} + \bar{\mathbf{S}}_{j,(i+1):(n-r)}\mathbf{z}_{(i+1):(n-r)}^{(k-1)}}{\bar{\mathbf{S}}_{j,i}}:
            \bar{\mathbf{S}}_{j,i} > 0, j=1,\ldots,n-r\Big\}, \\
        u_{i}^{(k)} &= \min\Big\{-\frac{\bar{\mathbf{S}}_{j,1:(i-1)}\mathbf{z}_{1:(i-1)}^{(k)} + \bar{\mathbf{S}}_{j,(i+1):(n-r)}\mathbf{z}_{(i+1):(n-r)}^{(k-1)}}{\bar{\mathbf{S}}_{j,i}}:
        \bar{\mathbf{S}}_{j,i} < 0, j=1,\ldots,n-r\Big\}
      \end{align*}
      \textnormal{with $l_{i}^{(k)} = -\infty$ ($u_{i}^{(k)} = \infty$) if $\bar{\mathbf{S}}_{j,i} > 0$ ($\bar{\mathbf{S}}_{j,i} < 0$) does not hold for any $j$.}
\item \textnormal{Compute $\bar{\mathbf{q}}_{1}^{(k)} = \mathbf{z}^{(k)}/\lVert \mathbf{z}^{(k)} \rVert$ and $\mathbf{q}_{1}^{(k)} = \mathbf{K}\mathbf{M}'\bar{\mathbf{q}}_{1}^{(k)}$.}
\end{list}
\end{center}
\end{algorithm}

After discarding an appropriate number of initial draws, $\mathbf{q}_{1}^{(k)}$ can be considered as dependent draws from the uniform distribution over $\mathcal{Q}_{1}(\bm{\phi}|\mathbf{F},\mathbf{S})$. To obtain (approximately) independent draws, keep only every $f$th draw, where $f$ is chosen such that so the retained draws are serially uncorrelated. To implement Step~1 in practice, I follow AD21 by drawing from the truncated standard normal distribution using the inverse cumulative distribution function (CDF) method. Letting $u \sim U(0,1)$, $\Phi^{-1}(u(\Phi(b) - \Phi(a)) + \Phi(a))$ is a truncated standard normal random variable with lower truncation point $a$ and upper truncation point $b$, where $\Phi(.)$ is the CDF of a standard normal random variable and $\Phi^{-1}(.)$ is the inverse CDF.

\section{Empirical Illustration and Comparison Against Alternative Algorithms}
\label{sec:empirical}

\setcounter{equation}{0}

This section applies the new algorithms in an empirical setting and compares their performance against existing alternatives.\footnote{All results are obtained using MATLAB 2020b on a laptop with Windows 10, an Intel Core i7-6700HQ CPU @ 2.60 GHz with four cores, and 8 GB of RAM.} The empirical application is from ACR19, who estimate the effects of monetary policy shocks in the United States.

\medskip

\noindent\textbf{Reduced-form VAR.} The model's endogenous variables are real GDP ($GDP_t$), the GDP deflator ($GDPDEF_t$), a commodity price index ($COM_t$), total reserves ($TR_t$), nonborrowed reserves ($NBR_t)$ (all in natural logarithms) and the federal funds rate ($FFR_t$). The data are monthly and run from January 1965 to June 2007. The VAR includes 12 lags.

I follow ACR19 by assuming a diffuse normal-inverse-Wishart prior over the reduced-form parameters. The posterior for the reduced-form parameters is then also a normal-inverse-Wishart distribution, from which it is straightforward to obtain independent draws (e.g. \citealp{DelNegro_Schorfheide_2011}).

\medskip

\noindent\textbf{Identifying restrictions.} Let $\mathbf{y}_{t} = (FFR_t,GDP_t,GDPDEF_t,COM_t,TR_t,NBR_t)'$. The monetary policy shock is $\varepsilon_{1t}$ and the first equation of the SVAR can be interpreted as the monetary policy reaction function. ACR19 set-identify the monetary policy shock using a mixture of sign and zero restrictions on the monetary policy reaction function. The zero restrictions are that $FFR_t$ does not react contemporaneously to $TR_t$ or $NBR_t$, which implies $\mathbf{e}_{1,6}'\mathbf{A}_{0}\mathbf{e}_{5,6} = (\bm{\Sigma}_{tr}^{-1}\mathbf{e}_{5,6})'\mathbf{q}_{1} = 0$ and $\mathbf{e}_{1,6}'\mathbf{A}_{0}\mathbf{e}_{6,6} = (\bm{\Sigma}_{tr}^{-1}\mathbf{e}_{6,6})'\mathbf{q}_{1} = 0$. The matrix containing the coefficients of the zero restrictions is $\mathbf{F}(\bm{\phi}) = (\bm{\Sigma}_{tr}^{-1}\mathbf{e}_{5,6}, \bm{\Sigma}_{tr}^{-1}\mathbf{e}_{6,6})'$. The sign restrictions are that, all else equal, $FFR_t$ is not decreased in response to higher $GDP_t$ or $GDPDEF_t$, which -- given the sign normalization $\mathbf{e}_{1,6}'\mathbf{A}_{0}\mathbf{e}_{1,6} = (\bm{\Sigma}_{tr}^{-1}\mathbf{e}_{1,6})'\mathbf{q}_{1} \geq 0$ -- implies $\mathbf{e}_{1,6}'\mathbf{A}_{0}\mathbf{e}_{2,6} = (\bm{\Sigma}_{tr}^{-1}\mathbf{e}_{2,6})'\mathbf{q}_{1} \leq 0$ and $\mathbf{e}_{1,6}'\mathbf{A}_{0}\mathbf{e}_{3,6} = (\bm{\Sigma}_{tr}^{-1}\mathbf{e}_{3,6})'\mathbf{q}_{1} \leq 0$. The impact response of $FFR_t$ to the monetary policy shock is also restricted to be nonnegative, which requires $\mathbf{e}_{1,6}'\mathbf{A}_{0}^{-1}\mathbf{e}_{1,6} = \mathbf{e}_{1,6}'\bm{\Sigma}_{tr}\mathbf{q}_{1} \geq 0$. The matrix containing the coefficients of the sign restrictions is $\mathbf{S}(\bm{\phi}) = (\bm{\Sigma}_{tr}^{-1}\mathbf{e}_{1,6}, -\bm{\Sigma}_{tr}^{-1}\mathbf{e}_{2,6}, -\bm{\Sigma}_{tr}^{-1}\mathbf{e}_{3,6}, (\mathbf{e}_{1,6}'\bm{\Sigma}_{tr})')'$.

I also consider other sets of identifying restrictions that add additional sign restrictions to $\mathbf{S}(\bm{\phi})$. Specifically, I add the sign restrictions on impulse responses proposed in \cite{Uhlig_2005}. These restrictions are that the impulse response of $FFR_t$ to the monetary policy shock is nonnegative for $h= 0,1,\ldots,H$ and the impulse responses of $GDPDEF_t$, $COM_t$ and $NBR_t$ are nonpositive for $h=0,1,\ldots,H$, where $H$ is a specified horizon. To explore how the algorithms perform under different numbers of sign restrictions, I consider $H \in \left\{5, 11, 23\right\}$. In total, there are 27~sign restrictions when $H = 5$, $51$~sign restrictions when $H=11$ and 99~sign restrictions when $H=23$.

\medskip

\noindent\textbf{Determining emptiness of $\mathcal{Q}_{1}(\bm{\phi}|\mathbf{F},\mathbf{S})$.} Given 1,000 draws from the posterior of $\bm{\phi}$, I check whether the identified set is empty using Algorithm~\ref{alg:checkis} and two alternative approaches. I compare the accuracy and speed of the three algorithms.

The first alternative is a rejection-sampling (RS) approach similar to that used by \cite{Arias_Rubio-Ramirez_Waggoner_2018} and \cite{Giacomini_Kitagawa_2021} to draw values of $\mathbf{Q}$. The algorithm draws $\mathbf{q}_{1}$ from a uniform distribution over $\mathcal{Q}_{1}(\bm{\phi}|\mathbf{F}) = \left\{\mathbf{q} \in \mathbb{S}^{n-1} : \mathbf{F}(\bm{\phi})\mathbf{q}_{1} = \mathbf{0}_{r\times 1}\right\}$ and checks whether the draw satisfies the sign restrictions.\footnote{The algorithm draws $\mathbf{z} \sim N(\mathbf{0}_{n\times 1},\mathbf{I}_n)$ and computes $\tilde{\mathbf{q}}_{1} = \left[\mathbf{I}_{n}-\mathbf{F}_{1}'(\mathbf{F}_{1}\mathbf{F}_{1}')^{-1}\mathbf{F}_{1}\right]\mathbf{z}$, so $\tilde{\mathbf{q}}_{1}$ satisfies the zero restrictions. $\tilde{\mathbf{q}}_{1}$ is then normalized so that it satisfies the sign normalization and has unit length before checking whether it satisfies the remaining sign restrictions.} If no draws of $\mathbf{q}_{1}$ satisfy the sign restrictions after $100,000$ draws, I approximate the identified set as empty. This algorithm may incorrectly classify the identified set as being empty, particularly when draws satisfying the zero restrictions satisfy the sign restrictions with low probability.

The second algorithm is from \nocite{Giacomini_Kitagawa_Volpicella_2021}Giacomini, Kitagawa and Volpicella (forthcoming). Their algorithm relies on the fact that any nonempty identified set for $\mathbf{q}_{1}$ must contain a vertex on the unit sphere where at least $n-1$ restrictions are binding. The algorithm determines whether the identified set is nonempty by considering all possible combinations of $n-r-1$ binding sign restrictions and checking whether the implied vertex satisfies the remaining sign restrictions. This approach will exactly determine whether the identified set is empty, but may become computationally burdensome when the number of sign restrictions is large, since it requires checking $\binom{s}{n-r-1}$ combinations of restrictions before concluding the identified set is empty.

\begin{table}[h]
    \caption{\label{tab:EmptyIS}Determining Emptiness of $\mathcal{Q}_{1}(\bm{\phi}|\mathbf{F},\mathbf{S})$}
    \begin{center}
     \begin{tabular*}{0.8\textwidth}{@{}crrrcrrr@{}}
     \hline\hline
     & \multicolumn{3}{c}{$\mathrm{Pr}(\mathcal{Q}_{1}(\bm{\phi}|\mathbf{F},\mathbf{S}) = \emptyset)$ (\%)} & \phantom{a} & \multicolumn{3}{c}{Computing Time (s)} \\ 
     Restrictions & A4.1 & RS & GKV & & A4.1 & RS & GKV \\ \hline
     (1) & 0.00 & 0.00 & 0.00 & & 9.86 & 0.04 & 0.17 \\
     (2) & 0.60 & 1.00 & 0.60 & & 9.26 & 6.55 & 4.26 \\
     (3) & 6.50 & 8.00 & 6.50  & & 9.38 & 50.55 & 103.78 \\
     (4) & 31.60 & 35.40 & 31.60  & & 9.64 & 223.51 & 3,277.2 \\     \hline\hline
     \end{tabular*}
    \end{center}
     \footnotesize\renewcommand{\baselineskip}{11pt}
     \textbf{Note:} (1) are the restrictions from ACR19 (2~zero restrictions, 4~sign restrictions); (2), (3) and (4) are the restrictions from ACR19 plus the restrictions from Uhlig (2005) with $H=5$ (27~sign restrictions), $H=11$ ($51$~sign restrictions) and $H=23$ (99~sign restrictions), respectively; A4.1 refers to Algorithm~4.1; RS refers to the rejection-sampling approach; GKV refers to the algorithm from Giacomini, Kitagawa and Volpicella (forthcoming).
\end{table}

To compare the accuracy of the algorithms, I compute the posterior probability that the identified set is empty. To compare computational efficiency, I tabulate the time taken to check whether the identified set is empty. Under the restrictions from ACR19, all three algorithms correctly determine that the identified set is nonempty at every draw of $\bm{\phi}$ (Table~\ref{tab:EmptyIS}). Although Algorithm~\ref{alg:checkis} is slower than the two alternatives in this case, in practice it would not be necessary to numerically check whether the identified set is nonempty, because the identified set is never empty when $r+s \leq n$.\footnote{This can be shown by applying Gordan's Theorem (e.g. p. 31 of \cite{Mangasarian_1969}) after transforming the system of identifying restrictions into a system of sign restrictions in a lower-dimensional space.} As the number of restrictions increases, the rejection-sampling approach misclassifies the identified set as being empty at some draws of $\bm{\phi}$. Algorithm~\ref{alg:checkis} is somewhat slower than the two alternatives under the second set of restrictions, but is much faster under the two larger sets of restrictions.

\medskip

\noindent\textbf{Obtaining uniform draws over $\mathcal{Q}_{1}(\bm{\phi}|\mathbf{F},\mathbf{S})$.} I verify that the Gibbs sampler generates draws from the uniform distribution over $\mathcal{Q}_{1}(\bm{\phi}|\mathbf{F},\mathbf{S})$ by comparing the distributions of draws obtained using the Gibbs sampler and the rejection sampler. Experiments using parallel Markov chains initialized at the same value suggest that dropping the first three draws from the Gibbs sampler is sufficient to remove dependence on the initial value; the distribution of the fourth draw across the parallel chains is statistically indistinguishable from the distribution generated by the rejection sampler (which generates independent draws). Dropping every second draw from the Gibbs sampler is sufficient to eliminate a significant first-order autocorrelation in the original set of draws.

\begin{figure}[h]
    \centering
    \caption{Histogram of Impulse Response Under Alternative Sampling Algorithms} \label{fig:histograms}
        \includegraphics[scale=0.5]{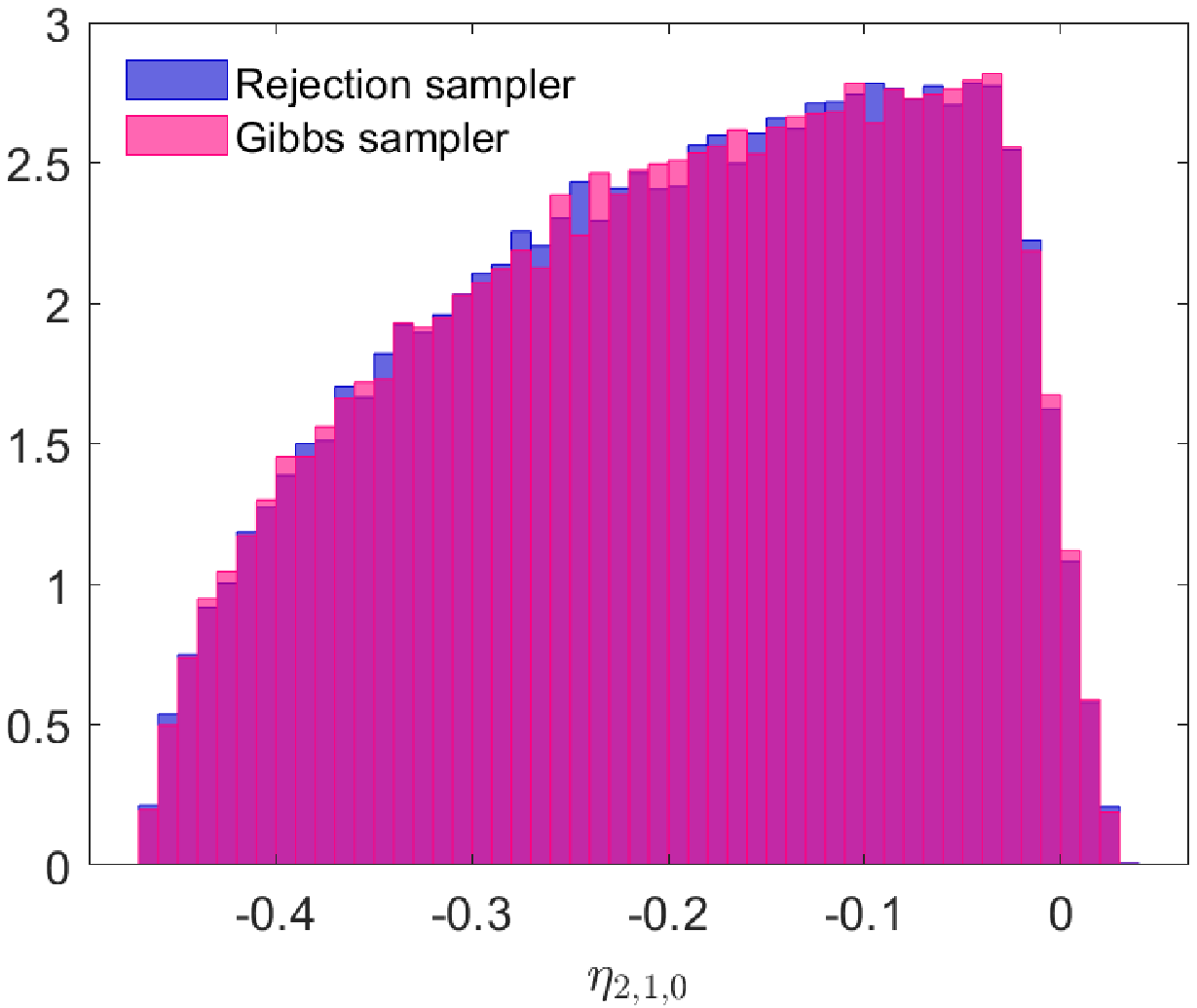} \\
    \begin{minipage}{\textwidth}
    \footnotesize\textbf{Note:} Impact response of output; based on 100,000 draws of $\mathbf{q}_{1}$ at random draw of $\bm{\phi}$.
    \end{minipage}
\end{figure}

Under the restrictions from ACR19 and given a single (random) draw of $\bm{\phi}$ with nonempty identified set, I obtain 100,000 draws of $\mathbf{q}$ from the Gibbs sampler after dropping the initial three draws and keeping every second draw. Figure~\ref{fig:histograms} plots histograms of the impact impulse response of output obtained using the two samplers; the distributions appear very similar and a two-sample Kolmogorov-Smirnov test fails to reject the null hypothesis that the two sets of draws are generated by the same distribution ($\text{p-value}=0.8$).

Next, I compare the efficiency of the two algorithms. I embed the Gibbs sampler (with a burn-in of three draws) and the rejection sampler within a standard posterior sampler for $\bm{\phi}$ to obtain 1,000 draws from the joint posterior of $(\bm{\phi},\mathbf{q}_{1})$ such that $\mathcal{Q}_{1}(\bm{\phi}|\mathbf{F},\mathbf{S})$ is nonempty. The rejection sampler is more efficient than the Gibbs sampler when there are few sign restrictions (Table~\ref{tab:samplingspeed}). The algorithms perform similarly when there is an intermediate number of sign restrictions. Under the larger sets of restrictions, the Gibbs sampler is more efficient.

\begin{table}[h]
    \caption{\label{tab:samplingspeed}Time Taken to Obtain 1,000 Draws (s)}
    \begin{center}
     \begin{tabular*}{0.7\textwidth}{@{}crrcrrr@{}}
     \hline\hline
        & \multicolumn{2}{c}{Draws of $(\bm{\phi},\mathbf{q}_{1})$} & & \multicolumn{3}{c}{Draws of $IS_{\eta}(\bm{\phi}|F,S)$} \\ \cmidrule{2-3} \cmidrule{5-7}
        Restrictions & Gibbs & RS & &  A4.1 & RS & GMM18 \\ \hline
        (1) & 13 & 3 & & 407 & 396 & 10 \\
        (2) & 13 & 13 & &  591 & 609 & 1537 \\
        (3) & 14 & 61 & & 759 & 797 & 10,638 \\
        (4) &  19  &  352   & &   887  &  1,270    &   85,990 \\
     \hline\hline
     \end{tabular*}
    \end{center}
     \footnotesize\renewcommand{\baselineskip}{11pt}
     \textbf{Note:} (1) are the restrictions from ACR19 (2~zero restrictions, 4~sign restrictions); (2), (3) and (4) are the restrictions from ACR19 plus the restrictions from Uhlig (2005) with $H=5$ (27~sign restrictions), $H=11$ ($51$~sign restrictions) and $H=23$ (99~sign restrictions), respectively; `Gibbs' refers to Algorithm~4.2 with a burn-in of three draws; RS refers to the rejection-sampling approach; GMM18 refers to the active-set algorithm from \cite{Gafarov_Meier_Montiel-Olea_2018}.
\end{table}

\medskip

\noindent\textbf{Computing the bounds of $IS_{\eta}(\bm{\phi}|\mathbf{F},\mathbf{S})$.} At 1,000 draws of $\bm{\phi}$ where $\mathcal{Q}_{1}(\bm{\phi}|\mathbf{F},\mathbf{S})$ is nonempty, I compute the lower and upper bounds of $IS_{\eta}(\bm{\phi}|\mathbf{F},\mathbf{S})$ when $\eta = \eta_{i,j,h}(\bm{\phi},\mathbf{q}_{1})$ is the output response to the monetary policy shock at horizons $h=0,\ldots,60$. The upper bound, $u(\bm{\phi})$, is defined as the value function of the optimization problem $u(\bm{\phi})=\max_{\mathbf{q}_{1} \in \mathcal{Q}_{1}(\bm{\phi}|\mathbf{F},\mathbf{S})}\eta_{i,j,h}(\bm{\phi},\mathbf{q}_{1})$, which is a quadratically constrained linear program with linear equality and inequality constraints. The lower bound, $l(\bm{\phi})$, is defined as the value function from the corresponding minimization problem.

I consider three alternative approaches. The first uses Algorithm~4.1 to check whether the identified set is nonempty and to obtain a value of $\mathbf{q}_{1}$ satisfying the identifying restrictions, which is used to initialize a gradient-based numerical optimization routine.\footnote{I use the interior-point algorithm in MATLAB's fmincon optimizer with analytical gradients of the objective function and constraints.} The second algorithm uses the same numerical optimization routine, but uses the rejection-sampling approach to obtain the initial value of $\mathbf{q}_{1}$. The third approach uses the active-set algorithm described in \cite{Gafarov_Meier_Montiel-Olea_2018}.\footnote{Given a set of binding restrictions, \citet{Gafarov_Meier_Montiel-Olea_2018} derive an expression for the value function (up to sign) of the optimization problems that define the bounds of the identified set. They also provide expressions for the corresponding solutions of the problems. They propose computing the value function at every possible combination of binding restrictions and checking whether the corresponding solution satisfies the remaining sign restrictions. The lower and upper bounds of the identified set are then the minimum and maximum, respectively, over the feasible value functions obtained at each combination of binding restrictions.} This approach may be computationally burdensome when the number of sign restrictions is large, since it requires computing the bounds of the identified set at $\sum_{k=0}^{n-r-1}\binom{s}{k}$ combinations of binding sign restrictions.

When there is a small number of sign restrictions, the algorithm from \cite{Gafarov_Meier_Montiel-Olea_2018} is the most efficient of the three (Table~\ref{tab:samplingspeed}). The other two algorithms perform similarly, since the bulk of the computing time is spent on the optimization step -- which is common across the two approaches -- rather than on trying to find a feasible initial value. As the number of sign restrictions increases, the algorithm from \cite{Gafarov_Meier_Montiel-Olea_2018} becomes computationally burdensome due to the explosion in the number of combinations of active restrictions to check. When $H=23$, using Algorithm~4.1 to obtain a feasible initial value for the numerical optimization routine is about 30~per cent faster than obtaining the initial value via rejection sampling.

\begin{figure}[h]
    \centering
    \caption{Output Response to a Monetary Policy Shock} \label{fig:outputresponses}
    \begin{tabular}{cc}
        \includegraphics[scale=0.4]{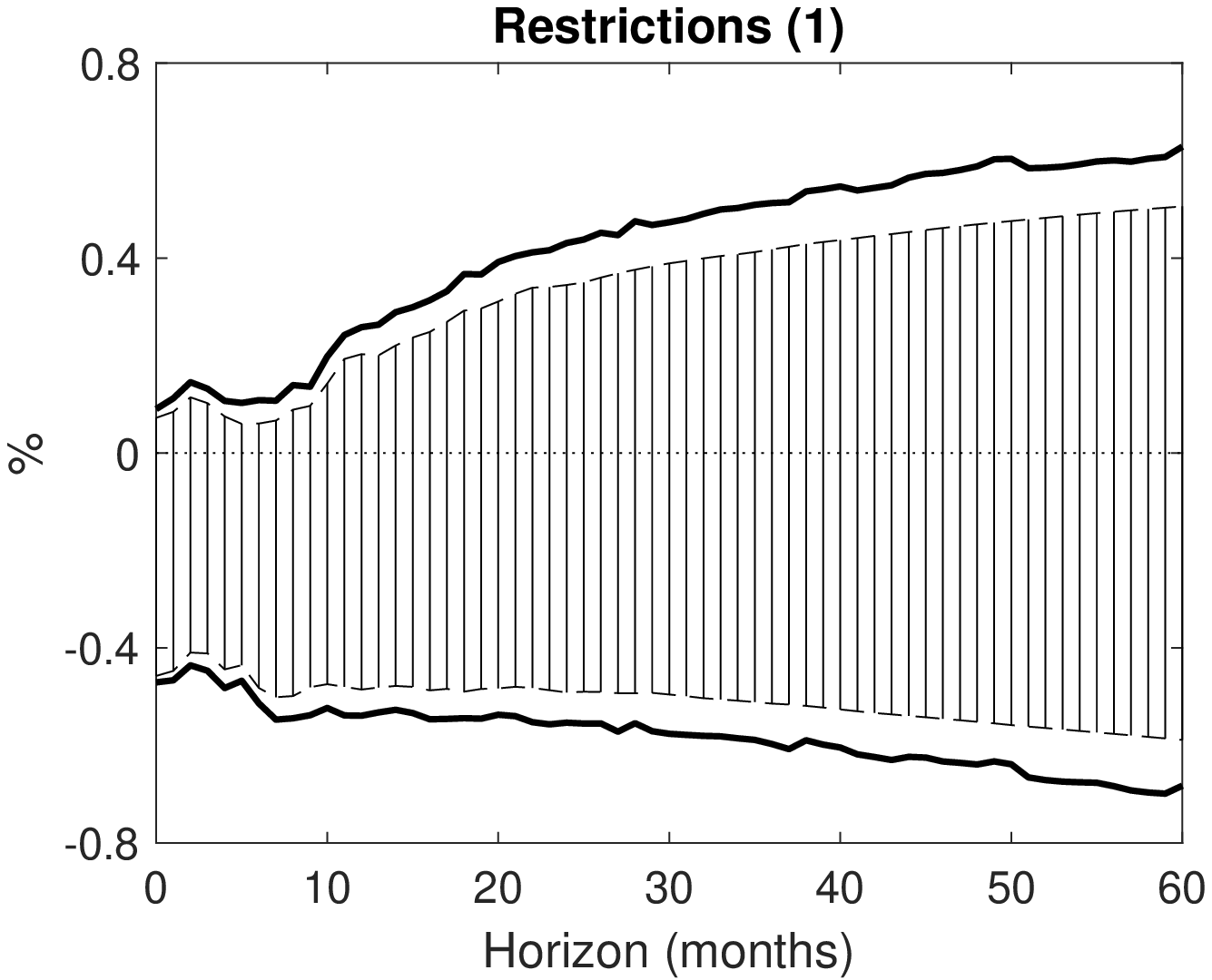} & \includegraphics[scale=0.4]{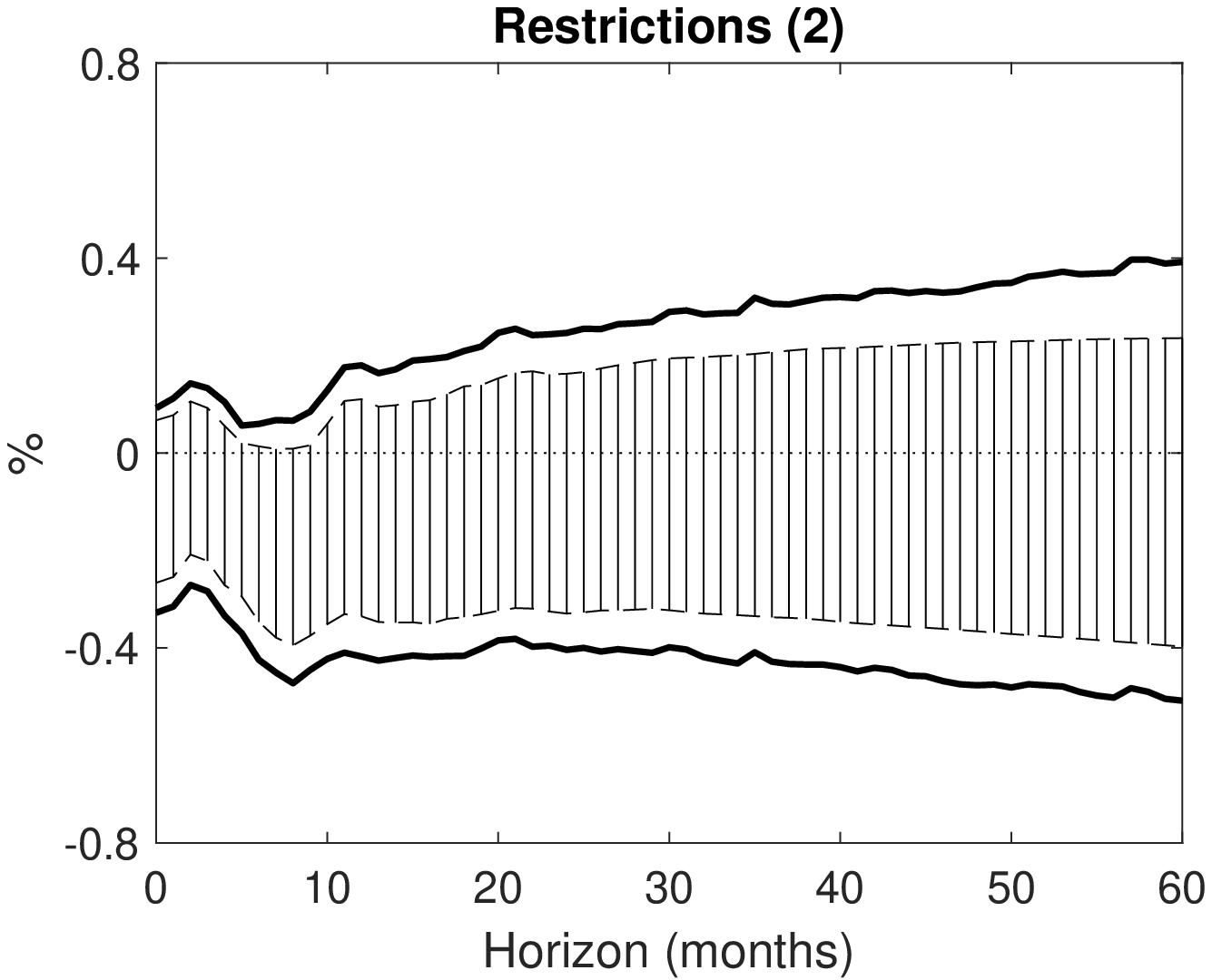} \\
        \includegraphics[scale=0.4]{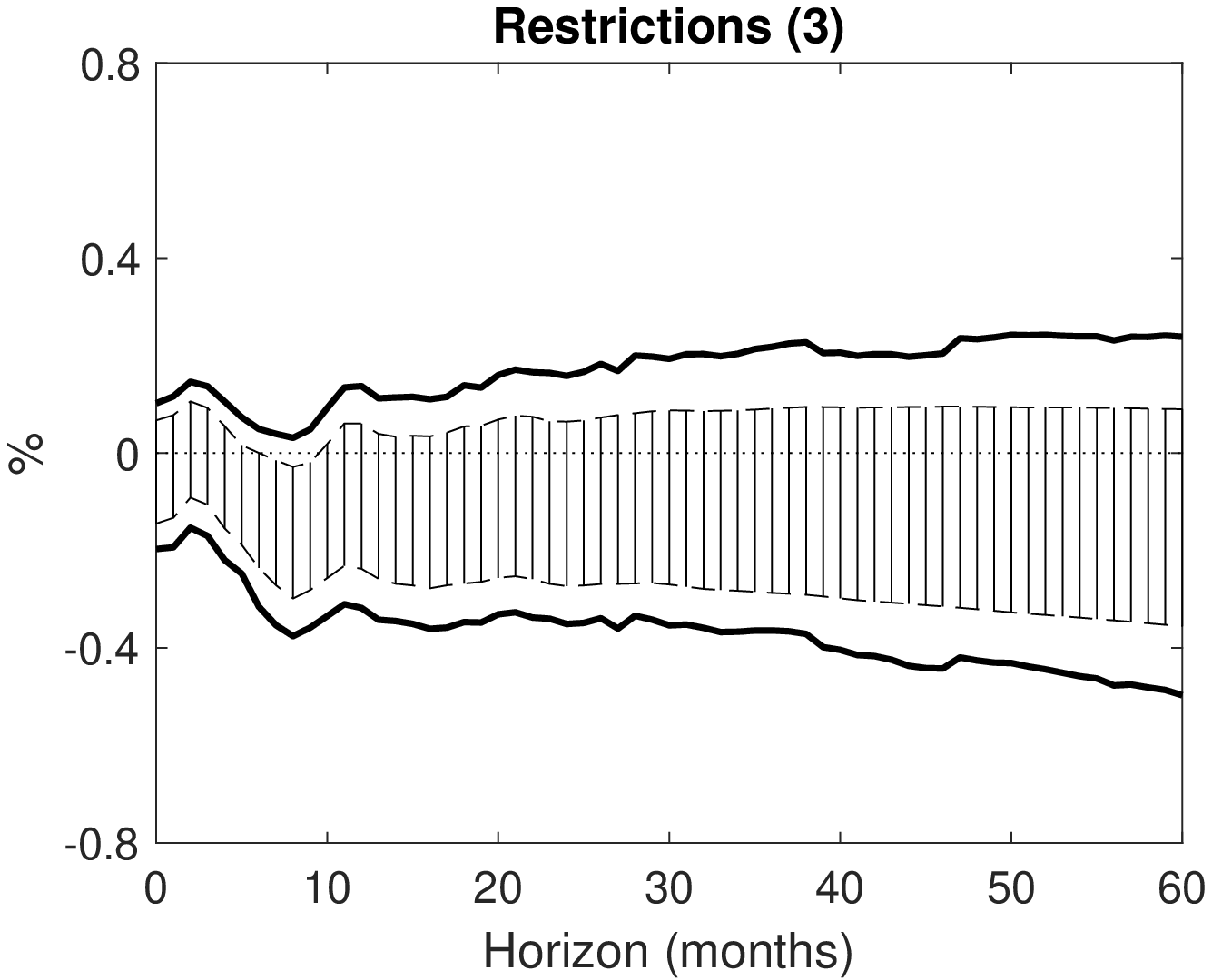} & \includegraphics[scale=0.4]{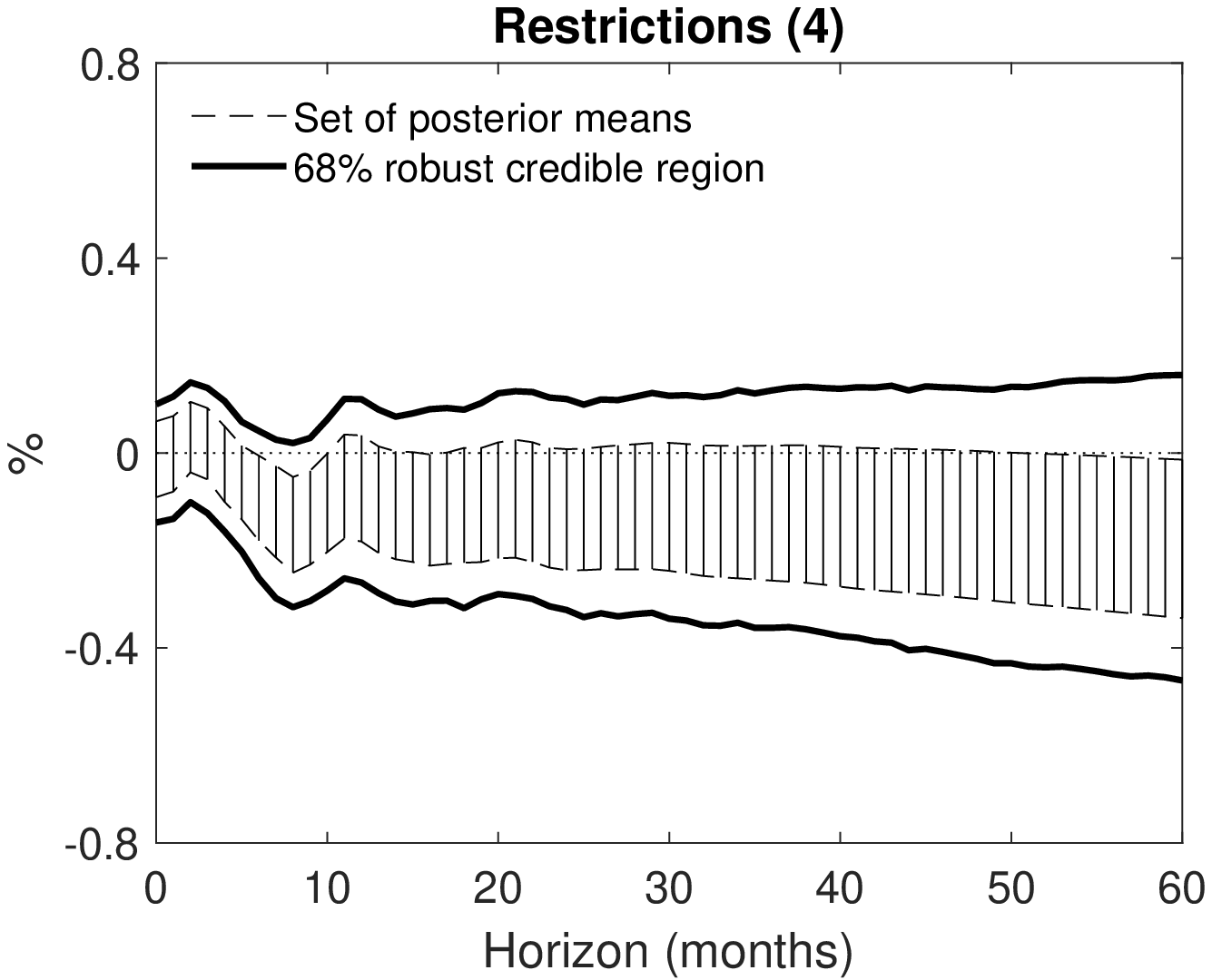} \\
    \end{tabular}
    \begin{minipage}{\textwidth}
    \footnotesize\textbf{Note:} (1) are the restrictions from ACR19 (2~zero restrictions, 4~sign restrictions); (2), (3) and (4) are the restrictions from ACR19 plus the restrictions from Uhlig (2005) with $H=5$ (27~sign restrictions), $H=11$ ($51$~sign restrictions) and $H=23$ (99~sign restrictions), respectively; responses are to a positive standard-deviation shock to the federal funds rate and are obtained using Algorithm~4.1 and a gradient-based numerical optimization routine.
    \end{minipage}
\end{figure}

To provide some sense of how tight the identified set is on average under the different identifying restrictions, Figure~\ref{fig:outputresponses} plots the set of posterior means and 68~per cent robust credible regions for the output response. These quantities are proposed by \cite{Giacomini_Kitagawa_2021} to assess or eliminate the sensitivity of posterior inference in set-identified models to the choice of conditional prior for $\mathbf{Q}|\bm{\phi}$. The set of posterior means is the average of $IS_{\eta}(\bm{\phi}|\mathbf{F},\mathbf{S})$ over the posterior for $\bm{\phi}$ and can be interpreted as a consistent estimator of the identified set. The robust credible region is the shortest interval covering 68~per cent of the posterior distribution under all possible conditional priors for $\mathbf{Q}|\bm{\phi}$ that satisfy the identifying restrictions, and can be interpreted as an asymptotically valid frequentist confidence interval. Each additional set of sign restrictions appears to appreciably truncate the identified set, on average. This explains the improvement in the performance of the proposed algorithms relative to those based on rejection sampling as the number of restrictions increases.

\medskip

\noindent\textbf{An extremely large number of restrictions.} This section provides an example of a set of identifying restrictions under which (accurate) posterior inference would be extremely computationally burdensome using existing algorithms. Specifically, in addition to the restrictions in ACR19 and the restrictions in \citet{Uhlig_2005} when $H=5$, I impose a narrative restriction on the timing of the maximum realisation of the monetary policy shock, as in \citet{Giacomini_Kitagawa_Read_2021b}. This `shock-rank' restriction is that the monetary policy shock in October 1979 -- the month in which Paul Volcker dramatically and unexpectedly raised the federal funds rate -- was the largest positive realization of the monetary policy shock in the sample. This restriction requires that $\varepsilon_{1k} = \mathbf{e}_{1,n}'\mathbf{A}_{0}\mathbf{u}_{k} = (\bm{\Sigma}_{tr}^{-1}\mathbf{u}_{k})'\mathbf{q}_{1} \geq 0$ and $\varepsilon_{1k} \geq \max_{t \neq k}\{\varepsilon_{1t}\}$ where $k$ is the index corresponding to October 1979. The latter restriction is equivalent to $(\bm{\Sigma}_{tr}^{-1}(\mathbf{u}_{k} - \mathbf{u}_{t}))'\mathbf{q}_{1} \geq 0$ for $t \neq k$. The restriction generates $T$ additional inequality restrictions on $\mathbf{q}_{1}$, where $T$ is the sample size and the inequality restrictions depend on the data (via the reduced-form VAR innovations).

\begin{figure}[h]
    \centering
    \caption{Impulse Responses to Monetary Policy Shock -- Shock-rank Restriction} \label{fig:shockrank}
    \begin{tabular}{cc}
        \includegraphics[scale=0.4]{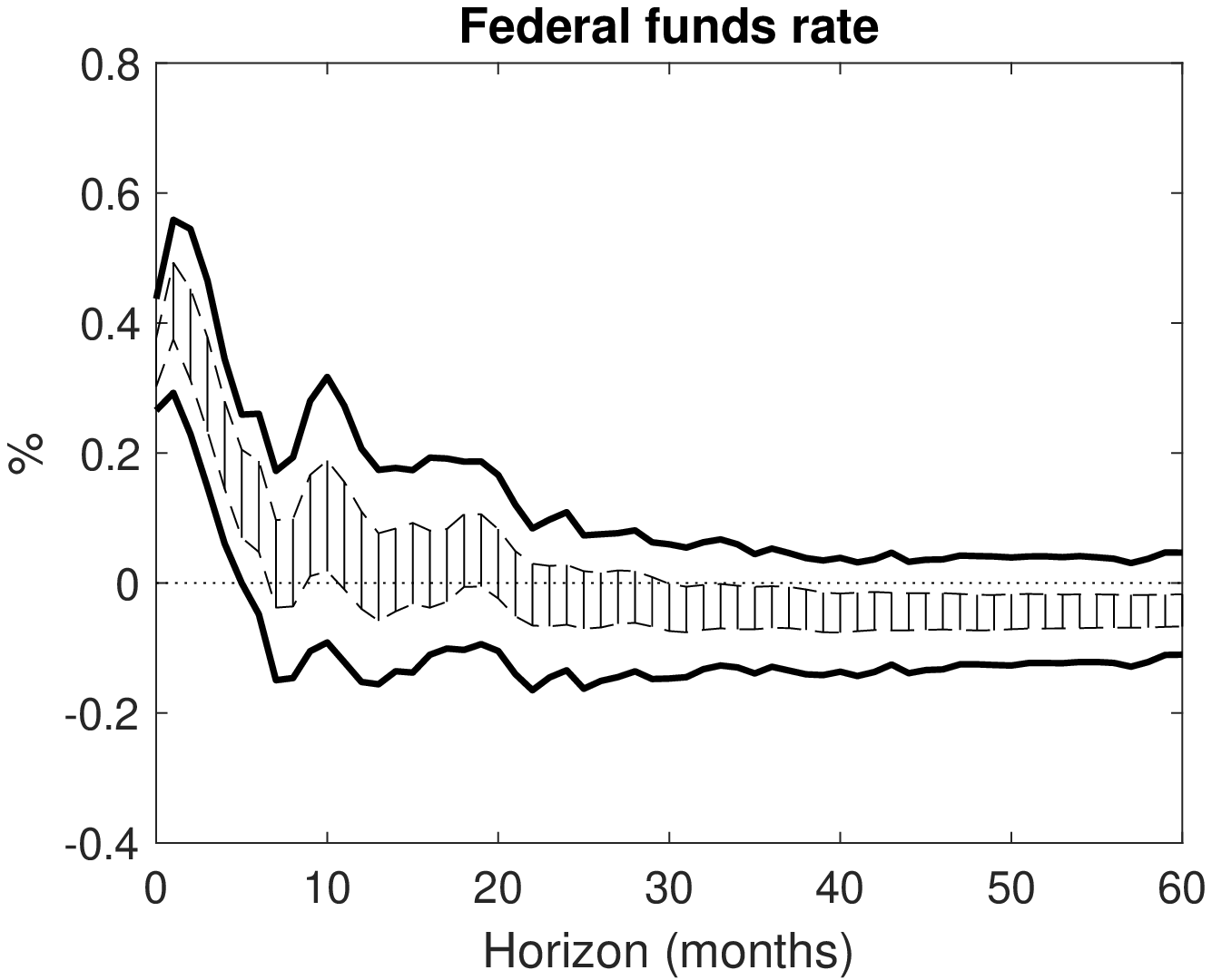} & \includegraphics[scale=0.4]{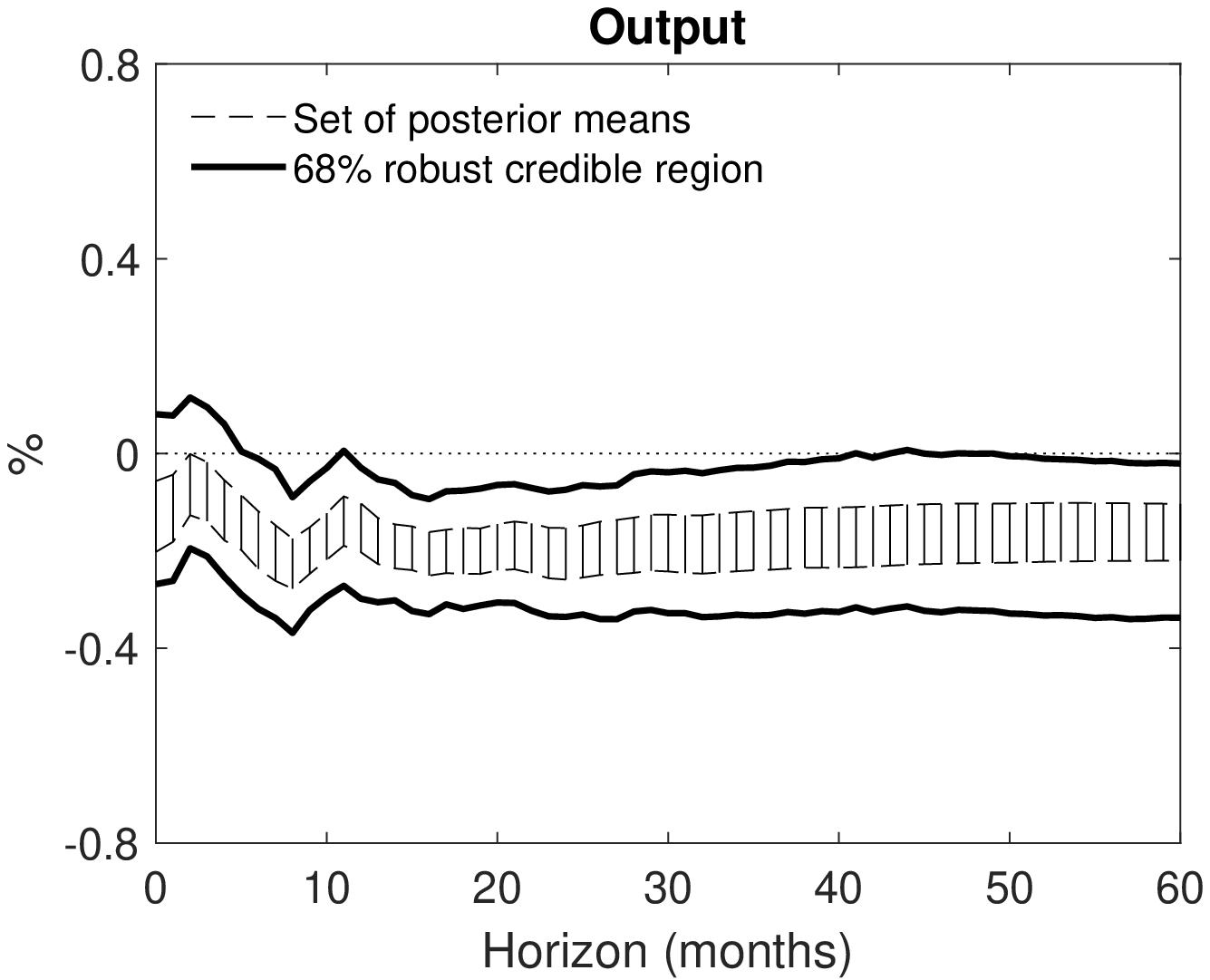} \\
    \end{tabular}
    \begin{minipage}{\textwidth}
    \footnotesize\textbf{Note:} Shock-rank restriction is the restriction that the monetary policy shock in October 1979 was the largest positive realization of the shock in the sample; this restriction is in addition to the restrictions from ACR and Uhlig (2005) with $H=5$ ($525$~sign restrictions in total); responses are to a positive standard-deviation shock to the federal funds rate and are obtained using Algorithm~4.1 and a numerical optimization procedure.
    \end{minipage}
\end{figure}

The restriction noticeably tightens the set of posterior means and robust credible intervals for the output response; the set of posterior means excludes zero at all horizons considered and the robust credible intervals exclude zero at most horizons (Figure~\ref{fig:shockrank}). The posterior lower probability -- the smallest probability attainable in the class of posteriors -- of a negative output response at the two-year horizon is around 95~per cent. The restriction therefore appears to be extremely informative when combined with the restrictions from ACR19 and \citet{Uhlig_2005}. However, the posterior probability that the identified set is empty is very high (around 96~per cent), which suggests that the restriction is inconsistent with the data.

The large system of inequality restrictions and narrow identified set generated by the restriction poses difficulties for existing algorithms. The approach in GKV would require checking $\binom{498+3+24}{3}= 23,979,550$ combinations of sign restrictions to determine that the identified set is empty. The approach in GMM would require considering $\sum_{k=0}^{3}\binom{498+3+24}{k}=24,117,626$ combinations of restrictions for every parameter of interest (i.e, for every variable and every impulse-response horizon) to compute the bounds of the identified set. Furthermore, given how tight the identified set appears to be on average over the posterior for $\bm{\phi}$, the rejection-sampling approach would require a very large number of draws of $\mathbf{q}_{1}$ given the zero restrictions to accurately approximate the posterior probability that the identified set is empty or to conduct inference under a conditionally uniform prior.

\medskip

\noindent\textbf{Discussion.} Overall, the results of this empirical exercise suggest that the new algorithms are likely to be preferable to existing alternatives when more than a handful of sign restrictions are imposed and/or these sign restrictions substantially truncate the identified set given the zero restrictions. The new algorithms should therefore facilitate the use of rich sets of sign restrictions alongside zero restrictions. It is difficult to provide definitive guidance about the number of restrictions above which the new algorithms will be more computationally efficient than the alternative algorithms considered; for example, whether the new algorithms are more efficient than the alternatives based on rejection sampling will depend on the extent to which the sign restrictions truncate the identified set given the zero restrictions. If practitioners want to use the fastest algorithm in any particular circumstance, they could test the algorithms against each other at particular values of the reduced-form parameters, such as at the maximum-likelihood estimate or at a small number of draws from the posterior.

\section{Extensions}
\label{sec:extensions}

\setcounter{equation}{0}

The algorithms described in the paper can be extended to some additional cases where there are restrictions on multiple columns of $\mathbf{Q}$.

\subsection{Some columns of $\mathbf{Q}$ are point-identified}

Consider the case where the first $i^{*}$ columns of $\mathbf{Q}$, $[\mathbf{q}_{1},\ldots,\mathbf{q}_{i^{*}}]$, are point-identified by zero restrictions, but interest is in impulse responses to the $j^{*} = (i^{*}+1)$th shock with both zero and sign restrictions on $\mathbf{q}_{j^{*}}$. Let $\mathbf{F}^{(i)}(\bm{\phi})$ represent the $r_{i} \times n$ matrix containing the coefficients of the zero restrictions constraining $\mathbf{q}_{i}$ and let
\begin{equation}
  \mathbf{F}(\bm{\phi},\mathbf{Q}) =
  \begin{bmatrix}
    \mathbf{F}^{(1)}(\bm{\phi})\mathbf{q}_{1} \\
    \vdots \\
    \mathbf{F}^{(n)}(\bm{\phi})\mathbf{q}_{n}
  \end{bmatrix}
  = \mathbf{0}_{\sum_{i=1}^{n}r_{i} \times 1}.
\end{equation}
If the zero restrictions do not constrain $\mathbf{q}_{i}$, then $\mathbf{F}^{(i)}(\bm{\phi})$ does not exist and $r_{i} = 0$. As in \cite{Giacomini_Kitagawa_2021}, assume the variables in $\mathbf{y}_{t}$ are ordered to satisfy the following ordering convention.
\begin{definition}
\textbf{Ordering convention.} Order the variables in $\mathbf{y}_{t}$ so that $r_{i}$ satisfies $r_{1} \geq r_{2} \geq \ldots \geq r_{n} \geq 0$. If the impulse response of interest is to the $j^{*}$th variable, order the $j^{*}$th variable first among ties.
\end{definition}

A necessary condition for exact identification of the first $i^{*}$ columns of $\mathbf{Q}$ is that $\mathrm{rank}(\mathbf{F}^{(i)}(\bm{\phi})) = r_{i} = n-i$ for $i = 1,\ldots,i^{*}$ (\cite{Rubio-Ramirez_Waggoner_Zha_2010}). A necessary and sufficient condition is that
\begin{equation}\label{eq:sufficient}
    \mathrm{rank}\left(
    \begin{bmatrix}
      \mathbf{F}^{(i)}(\bm{\phi})', \mathbf{q}_{1}, \ldots, \mathbf{q}_{i-1}
    \end{bmatrix}
    '\right) = n - 1, \quad \text{for $i=1,\ldots,i^{*}$},
\end{equation}
which additionally requires that the restrictions in $\mathbf{F}^{(i)}(\bm{\phi})$, $i=1,\ldots,i^{*}$, are `non-redundant' in the sense discussed in \cite{Bacchiocchi_Kitagawa_2021}.\footnote{If the condition in Equation~(\ref{eq:sufficient}) is satisfied, then $\mathbf{q}_{i}$ is restricted to lie in the one-dimensional linear subspace of $\mathbb{R}^{n}$ satisfying $(\mathbf{F}^{(i)}(\bm{\phi})', \mathbf{q}_{1}, \ldots, \mathbf{q}_{i-1})'\mathbf{q}_{i} = \mathbf{0}_{(n-1)\times 1}$. The sign normalization and requirement that $\mathbf{q}_{i}$ lie on the unit sphere pin down $\mathbf{q}_{i}$.}

Assume $r_{j^{*}} < n-j^{*}$, in which case $\mathbf{q}_{j^{*}}$ is set-identified. $\mathbf{q}_{j^{*}}$ is constrained by the sign restrictions $\mathbf{S}(\bm{\phi})\mathbf{q}_{j^{*}} \geq \mathbf{0}_{s\times 1}$ and an extended set of zero restrictions incorporating the restriction that $\mathbf{q}_{j^{*}}$ is orthogonal to the preceding columns of $\mathbf{Q}$:
\begin{equation}
      \ddot{\mathbf{F}}^{(j^{*})}(\bm{\phi})\mathbf{q}_{j^{*}} =
      \begin{bmatrix}
      \mathbf{F}^{(j^{*})}(\bm{\phi})', \mathbf{q}_{1}, \ldots, \mathbf{q}_{i^{*}}
    \end{bmatrix}
    '\mathbf{q}_{j^{*}} = \mathbf{0}_{(r_{j^{*}} + j^{*} - 1)\times 1},
\end{equation}
where I have suppressed the dependence of $\ddot{\mathbf{F}}^{(j^{*})}(\bm{\phi})$ on $\mathbf{F}^{(i)}(\bm{\phi})$, $i=1,\ldots,i^{*}$. If the conditions for exact identification above are satisfied, each $\mathbf{q}_{i}$, $i=1,\ldots,i^{*}$, can be determined iteratively as follows. First, find a unit-length vector $\mathbf{q}_{1}$ satisfying $\mathbf{F}^{(1)}(\bm{\phi})\mathbf{q}_{1} = \mathbf{0}_{(n-1)\times 1}$ by computing an orthonormal basis for the null space of $\mathbf{F}^{(1)}(\bm{\phi})$. Normalize $\mathbf{q}_{1}$ so it satisfies the sign normalization $(\bm{\Sigma}_{tr}^{-1}\mathbf{e}_{1,n})'\mathbf{q}_{1} \geq 0$. Then, for $i=2,\ldots,i^{*}$, find a unit-length vector $\mathbf{q}_{i}$ satisfying $(\mathbf{F}^{(i)}(\bm{\phi})',\mathbf{q}_{1},\ldots,\mathbf{q}_{i-1})'\mathbf{q}_{i} = \mathbf{0}_{(n-1)\times 1}$ by computing an orthonormal basis for the null space of $(\mathbf{F}^{(i)}(\bm{\phi})',\mathbf{q}_{1},\ldots,\mathbf{q}_{i-1})'$. Normalize $\mathbf{q}_{i}$ so it satisfies the sign normalization $(\bm{\Sigma}_{tr}^{-1}\mathbf{e}_{i,n})'\mathbf{q}_{i} \geq 0$. Under the assumption that $\mathrm{rank}(\ddot{\mathbf{F}}^{(j^{*})}(\bm{\phi})) = r_{j^{*}} + j^{*} - 1$, $\ddot{\mathbf{F}}^{(j^{*})}(\bm{\phi})$ can replace $\mathbf{F}(\bm{\phi})$ in the algorithms described in Section~\ref{sec:numerical} without further modification.

\subsection{A subset of the columns of $\mathbf{Q}$ is determined up to a linear subspace}

The algorithms can also be applied when $(\mathbf{q}_{1},\ldots,\mathbf{q}_{i^{*}})$ is not point-identified, but is pinned down to lie within an $i^{*}$-dimensional linear subspace of $\mathbb{R}^{n}$. Consider the case where $\mathbf{F}^{(1)}(\bm{\phi}) = \ldots = \mathbf{F}^{(i^{*})}(\bm{\phi})$, so $r_{1} = \ldots = r_{i^{*}} \equiv \tilde{r}$, and assume $\tilde{r} = n - i^{*}$, which implies $\mathbf{q}_{i}$ is set-identified for $i = 1,\ldots,i^{*}$. For example, this pattern of restrictions arises in proxy SVARs when there are multiple proxies for multiple shocks (e.g. \nocite{Giacomini_Kitagawa_Read_2021a}Giacomini, Kitagawa and Read, forthcoming). The restriction $\mathbf{F}^{(i)}(\bm{\phi})\mathbf{q}_{i} = \mathbf{0}_{\tilde{r} \times 1}$ restricts $\mathbf{q}_{i}$ to lie in a linear subspace of $\mathbb{R}^{n}$ with dimension $n-\tilde{r} = i^{*}$. Since $\mathbf{F}^{(i)}(\bm{\phi})$ is common for $i=1,\ldots,i^{*}$, the first $i^{*}$ columns of $\mathbf{Q}$ are restricted to lie in the same $i^{*}$-dimensional subspace. An orthonormal basis for this subspace is $N(\mathbf{F}^{(1)}(\bm{\phi}))$. $\mathbf{q}_{j}^{*}$ must be orthogonal to the preceding columns of $\mathbf{Q}$, so it must be orthogonal to the $i^{*}$-dimensional subspace spanned by the columns of $N(\mathbf{F}^{(1)}(\bm{\phi}))$. $\mathbf{q}_{j^{*}}$ must therefore satisfy $\mathbf{S}(\bm{\phi})\mathbf{q}_{j^{*}} \geq \mathbf{0}_{s\times 1}$ and the extended set of zero restrictions
\begin{equation}
      \ddot{\mathbf{F}}^{(j^{*})}(\bm{\phi})\mathbf{q}_{j^{*}} =
      \begin{bmatrix}
      \mathbf{F}^{(j^{*})}(\bm{\phi}) \\
      N(\mathbf{F}^{(1)}(\bm{\phi}))'
    \end{bmatrix}
    \mathbf{q}_{j^{*}} = \mathbf{0}_{(r_{j^{*}} + j^{*} - 1)\times 1}.
\end{equation}
Under the assumption $\mathrm{rank}(\ddot{\mathbf{F}}^{(j^{*})}(\bm{\phi})) = r_{j^{*}} + j^{*}-1 < n - j^{*}$, $\ddot{\mathbf{F}}^{(j^{*})}(\bm{\phi})$ can replace $\mathbf{F}(\bm{\phi})$ in the algorithms described in Section~\ref{sec:numerical} without further modification.

\subsection{Sign and zero restrictions on multiple columns of $\mathbf{Q}$}

Consider the case where there are are zero and sign restrictions on the first $i^{*}<n$ columns of $\mathbf{Q}$ with $0 \leq r_{i} < n-i$ for $i=1,\ldots,i^{*}$, so $\mathbf{q}_{i}$ is set-identified for all $i=1,\ldots,n$. Let $\mathbf{S}^{(i)}(\bm{\phi})\mathbf{q}_{i} \geq \mathbf{0}_{s_{i} \times 1}$ represent the sign restrictions constraining $\mathbf{q}_{i}$ and let $\mathbf{S}(\bm{\phi},\mathbf{Q}) \geq \mathbf{0}_{\sum_{i=1}^{n} s_{i} \times 1}$ represent the system of sign restrictions (including the sign normalizations). The identified set for $\mathbf{Q}$ is
\begin{equation}
  \mathcal{Q}(\bm{\phi}|\mathbf{F},\mathbf{S}) = \left\{\mathbf{Q} \in \mathcal{O}(n): \mathbf{F}(\bm{\phi},\mathbf{Q}) = \mathbf{0}_{\sum_{i=1}^{n} r_{i} \times 1}, \mathbf{S}(\bm{\phi},\mathbf{Q}) \geq \mathbf{0}_{\sum_{i=1}^{n} s_{i} \times 1}\right\}.
\end{equation}
When there are sign restrictions only, AD21 provide a sufficient condition for checking whether $\mathcal{Q}(\bm{\phi}|\mathbf{F},\mathbf{S})$ is empty and a sufficient condition for checking whether it is nonempty. These sufficient conditions can be extended to the case with zero restrictions in the following way.

First, a sufficient condition for $\mathcal{Q}(\bm{\phi}|\mathbf{F},\mathbf{S}) = \emptyset$ is $\mathcal{Q}_{i}(\bm{\phi}|\mathbf{F}^{(i)},\mathbf{S}^{(i)}) = \emptyset$ for any $i \in \{1,\ldots,i^{*}\}$.  Intuitively, if the identified set for a single column of $\mathbf{Q}$ is empty when imposing only the restrictions directly constraining that column, the identified set for $\mathbf{Q}$ itself must be empty. To check this, one can apply Algorithm~\ref{alg:checkis} for each $i$, replacing $\mathbf{F}(\bm{\phi})$ and $\mathbf{S}(\bm{\phi})$ with $\mathbf{F}^{(i)}(\bm{\phi})$ and $\mathbf{S}^{(i)}(\bm{\phi})$, respectively.

Second, a sufficient condition for $\mathcal{Q}(\bm{\phi}|\mathbf{F},\mathbf{S})$ to be nonempty is that $\mathcal{Q}_{1}(\bm{\phi}|\mathbf{F}^{(1)},\mathbf{S}^{(1)})$ is nonempty and $\mathcal{Q}_{i}(\bm{\phi}|\hat{\mathbf{F}}^{(i)},\mathbf{S}^{(i)})$ is nonempty for all $i=2,\ldots,i^{*}$, where
\begin{equation}
  \hat{\mathbf{F}}^{(i)}(\bm{\phi}) =
  \begin{bmatrix}
    \mathbf{F}^{(i)}(\bm{\phi})', \mathbf{q}^{(0)}_{1}, \ldots, \hat{\mathbf{q}}^{(0)}_{i-1}
  \end{bmatrix}
  '.
\end{equation}
$\mathbf{q}^{(0)}_{1}$ is the transformed Chebyshev centre obtained from applying Algorithm~\ref{alg:checkis} under the system of restrictions $\mathbf{F}^{(1)}(\bm{\phi})\mathbf{q}_{1} = \mathbf{0}_{r_{1}\times 1}$ and $\mathbf{S}^{(1)}(\bm{\phi})\mathbf{q}_{1} \geq \mathbf{0}_{s_{1}\times 1}$, and $\hat{\mathbf{q}}^{(0)}_{i}$ is the transformed Chebyshev centre obtained from applying Algorithm~\ref{alg:checkis} under the system of restrictions $\hat{\mathbf{F}}^{(i)}(\bm{\phi})\mathbf{q}_{i} = \mathbf{0}_{r_{i}\times 1}$ and $\mathbf{S}^{(i)}(\bm{\phi})\mathbf{q}_{i} \geq \mathbf{0}_{s_{i}\times 1}$. It is only necessary to check this condition for the first $i^{*}$ columns of $\mathbf{Q}$, since a set of $n-i^{*}$ orthonormal vectors satisfying the sign normalizations can always be constructed in the null space of $(\mathbf{q}_{1}^{(0)},\ldots,\mathbf{q}_{i^{*}}^{(0)})$.

If neither sufficient condition is satisfied, one could attempt to determine whether the identified set is nonempty using rejection sampling. However, as in the case where a single column of $\mathbf{Q}$ is restricted, this is likely to be inaccurate or computationally burdensome when the sign restrictions markedly tighten the identified set for $\mathbf{Q}$ given the zero restrictions.

AD21 also describe a Gibbs sampler that is applicable when there are sign restrictions on multiple columns of $\mathbf{Q}$. Similar to the case where a single column of $\mathbf{Q}$ is restricted, this sampler can be extended to allow for zero restrictions. Assume one is able to obtain a value of $\mathbf{Q}_{1:i^{*}} = (\mathbf{q}_{1},\ldots,\mathbf{q}_{i^{*}})$ satisfying the sign and zero restrictions and that the parameter of interest is a function of $\mathbf{Q}_{1:i^{*}}$ (e.g. an impulse response to one of the first $i^{*}$ shocks). Also, assume that $r_{i} < n-i^{*}$ for all $i=1,\ldots,n$ and that $i^{*} < n-1$.

\begin{algorithm} \label{alg:gibbsQ}
\textnormal{\textbf{Gibbs sampler for $\mathbf{Q}_{1:i^{*}}$.} Assume $\mathbf{Q}^{(0)}_{1:i^{*}} = (\mathbf{q}^{(0)}_{1},\ldots,\mathbf{q}^{(0)}_{i^{*}})$ is available satisfying the system of identifying restrictions. Let $L$ be the desired number of draws of $\mathbf{Q}_{1:i^{*}}$. For each $k=1,\ldots,L$, sequentially complete the following steps for $j=1,\ldots,i^{*}$:}
\setcounter{bean}{0}
\begin{center}
\begin{list}
{\textsc{Step} \arabic{bean}.}{\usecounter{bean}}
  \item \textnormal{Compute $\mathbf{F}^{\dagger} = (\mathbf{q}_{1}^{(k)},\ldots,\mathbf{q}_{j-1}^{(k)},\mathbf{q}_{j+1}^{(k-1)},\ldots,\mathbf{q}_{i^{*}}^{(k-1)},\mathbf{F}_{j}')'$.}
    \item \textnormal{Compute the change-of-basis matrix $\mathbf{K} = (N(\mathbf{F}^{\dagger}),N(N(\mathbf{F}^{\dagger})'))$ and transform the coefficient vectors of the sign, zero and orthogonality restrictions into the new basis via $\tilde{\mathbf{S}}_{j} = (\mathbf{K}^{-1}\mathbf{S}_{j}')'$ and $\tilde{\mathbf{F}}= (\mathbf{K}^{-1}\mathbf{F}^{\dagger\prime})'$.}
    \item \textnormal{Project the coefficient vectors of the sign restrictions in the new basis onto the linear subspace spanned by the rows of $\tilde{\mathbf{F}}$ and drop the last $r_{j}+i^{*}-1$ elements of the resulting vectors. The transformed matrix of coefficients is $\bar{\mathbf{S}}_{j} = (\mathbf{M}(\mathbf{I}_{n} - \tilde{\mathbf{F}}'(\tilde{\mathbf{F}}\tilde{\mathbf{F}}')^{-1}\tilde{\mathbf{F}})\tilde{\mathbf{S}}')'$, where  $\mathbf{M} = (\mathbf{I}_{n-r_{j}-i^{*}-1},\mathbf{0}_{(n-r_{j}-i^{*}-1)\times (r_{j}+i^{*}+1)})$.}
  \item \textnormal{Set $\mathbf{z}_{j}^{(k-1)} = \mathbf{M}\mathbf{K}^{-1}\mathbf{q}_{j}^{(k-1)}$ and apply Steps~1 and 2 of Algorithm~4.2 for $i=1,\ldots,n-r_{j}-i^{*}+1$ to obtain $\mathbf{q}_{j}^{(k)}$.}
\end{list}
\end{center}
\end{algorithm}

The key difference between this algorithm and Algorithm~4.2 is that, within every iteration of the Gibbs sampler, the columns of $\mathbf{Q}_{1:i^{*}}$ other than $\mathbf{q}_{j}$ are treated as given. The condition that $\mathbf{q}_{j}$ is orthogonal to the remaining columns of $\mathbf{Q}_{1:i^{*}}$ can be treated as a set of zero restrictions. An important assumption is that an initial value of $\mathbf{Q}_{1:i^{*}}$ satisfying the identifying restrictions is available; when the sufficient condition for a nonempty identified set described above is not satisfied, obtaining such a value may require the use of rejection-sampling methods. Numerical exercises indicate that this algorithm draws from the same uniform distribution over the identified set for $\mathbf{Q}_{1:i^{*}}$ as the rejection samplers described in \cite{Arias_Rubio-Ramirez_Waggoner_2018} and \cite{Giacomini_Kitagawa_2021}. The algorithm is likely to be more efficient than rejection sampling when sign restrictions considerably truncate the identified set for $\mathbf{Q}_{1:i^{*}}$ given the zero restrictions. The algorithm may be useful for approximating the bounds of the identified set for a scalar parameter of interest.\footnote{When multiple columns of $\mathbf{Q}$ are constrained by the identifying restrictions, the approach from \cite{Gafarov_Meier_Montiel-Olea_2018} to computing the bounds of the identified set is generally inapplicable. Moreover, computing these bounds via numerical optimization may be difficult, since the optimization problem is nonconvex. For a rejection-sampling approach to this problem, see Algorithm~2 of \cite{Giacomini_Kitagawa_2021}.}

\section{Conclusion}
\label{sec:conclusion}

\setcounter{equation}{0}

In SVAR models, a system of sign and zero restrictions constraining a single column of the orthonormal matrix can be expressed as a system of sign restrictions in a lower-dimensional space. Consequently, algorithms that are useful for conducting Bayesian inference under sign restrictions can be extended to the case where there are also zero restrictions. I show that such algorithms can be more accurate and computationally efficient than existing alternatives, particularly when a large number of sign restrictions considerably truncates the identified set given the zero restrictions. The algorithms in this paper should therefore facilitate Bayesian inference when rich sets of sign restrictions are imposed alongside zero restrictions.

\section*{Acknowledgements}
The author would like to thank Thorsten Drautzburg, Raffaella Giacomini, Vincent Sterk, two anonymous referees and the editor, Michael Jansson, for helpful feedback. The views expressed in this paper are those of the author and do not reflect the views of the Reserve Bank of Australia.

\bibliography{SVARAlgorithms_Bibliography}


    \section*{Appendix A: Proofs of Results}
    \renewcommand{\theequation}{A.\arabic{equation}}
    \renewcommand{\thesection}{A}
    \setcounter{equation}{0}

    \medskip

    \textbf{Proof of Proposition~\ref{prop:equivalence}:} (a) Assume $\mathbf{q}_{1} \in \mathbb{R}^{n}$ satisfies $\mathbf{F}(\bm{\phi})\mathbf{q}_{1} = \mathbf{0}_{r\times 1}$ and $\mathbf{S}(\bm{\phi})\mathbf{q}_{1} \geq \mathbf{0}_{s\times 1}$, and let $\tilde{\mathbf{q}}_{1} = \mathbf{K}^{-1}\mathbf{q}_{1}$. Since $\mathbf{K}$ is orthonormal, $\tilde{\mathbf{F}}(\bm{\phi})\tilde{\mathbf{q}}_{1} = (\mathbf{K}^{-1}\mathbf{F}(\bm{\phi})')'\mathbf{K}^{-1}\mathbf{q}_{1} = \mathbf{F}(\bm{\phi})\mathbf{q}_{1} = \mathbf{0}_{r\times 1}$ and $\tilde{\mathbf{S}}(\bm{\phi})\tilde{\mathbf{q}}_{1} = (\mathbf{K}^{-1}\mathbf{S}(\bm{\phi})')' \mathbf{K}^{-1}\mathbf{q}_{1} = \mathbf{S}(\bm{\phi})\mathbf{q}_{1} \geq \mathbf{0}_{s\times 1}$, so $\tilde{\mathbf{q}}_{1}$ satisfies the restrictions in the transformed basis. It follows that
\begin{align*}
  \bar{\mathbf{S}}_{i}(\bm{\phi})\tilde{\mathbf{q}}_{1} &= \left[\left(\mathbf{I}_{n} - \tilde{\mathbf{F}}(\bm{\phi})'\left(\tilde{\mathbf{F}}(\bm{\phi})\tilde{\mathbf{F}}(\bm{\phi})'\right)^{-1}\tilde{\mathbf{F}}(\bm{\phi})\right)\tilde{\mathbf{S}}_{i}(\bm{\phi})'\right]'\tilde{\mathbf{q}}_{1}\\
  &= \tilde{\mathbf{S}}_{i}(\bm{\phi})\tilde{\mathbf{q}}_{1} - \tilde{\mathbf{S}}_{i}(\bm{\phi})\tilde{\mathbf{F}}(\bm{\phi})'\left(\tilde{\mathbf{F}}(\bm{\phi})\tilde{\mathbf{F}}(\bm{\phi})'\right)^{-1}\tilde{\mathbf{F}}(\bm{\phi})\tilde{\mathbf{q}}_{1} \\
  &\geq 0,
\end{align*}
for $i=1,\ldots,s$, where the final line uses $\tilde{\mathbf{F}}(\bm{\phi})\tilde{\mathbf{q}}_{1} = \mathbf{0}_{r\times 1}$ and $\tilde{\mathbf{S}}_{i}(\bm{\phi})\tilde{\mathbf{q}}_{1} \geq 0$. $\tilde{\mathbf{q}}_{1}$ therefore satisfies the sign restrictions after projecting their coefficient vectors onto the hyperplane generated by the zero restrictions. Since $\bar{\mathbf{S}}_{i}(\bm{\phi})$ and $\tilde{\mathbf{q}}_{1}$ both lie in the hyperplane spanned by the first $n-r$ basis vectors, their last $r$ elements are equal to zero, so $\bar{\mathbf{S}}_{i}(\bm{\phi})\tilde{\mathbf{q}}_{1} = (\mathbf{M}\bar{\mathbf{S}}_{i}(\bm{\phi})')'\mathbf{M}\tilde{\mathbf{q}}_{1} \geq 0$. It follows that $\bar{\mathbf{S}}(\bm{\phi})\bar{\mathbf{q}}_{1} \geq \mathbf{0}_{s\times 1}$, where $\bar{\mathbf{S}}(\bm{\phi})' = \mathbf{M}(\bar{\mathbf{S}}_{i}(\bm{\phi})',\ldots,\bar{\mathbf{S}}_{s}(\bm{\phi})')$ and $\bar{\mathbf{q}}_{1} = \mathbf{M}\tilde{\mathbf{q}}_{1} = \mathbf{M}\mathbf{K}^{-1}\mathbf{q}_{1}$.

(b) Assume that $\bar{\mathbf{q}}_{1} \in \mathbb{R}^{n-r}$ satisfies $\bar{\mathbf{S}}(\bm{\phi})\bar{\mathbf{q}}_{1} \geq \mathbf{0}_{s\times 1}$. Given the definition of $\mathbf{M}$, $\mathbf{M}'$ is the $n\times (n-r)$ matrix such that, for an $(n-r)\times 1$ vector $\mathbf{x}$, $\mathbf{M}'\mathbf{x} = (\mathbf{x}',\mathbf{0}_{1\times r})'$. It follows that $(\mathbf{M}'\bar{\mathbf{S}}(\bm{\phi})')'\mathbf{M}'\bar{\mathbf{q}}_{1} \geq \mathbf{0}_{s\times 1}$. The $i$th column of $\mathbf{M}'\bar{\mathbf{S}}(\bm{\phi})'$ is $\bar{\mathbf{S}}_{i}(\bm{\phi})'$ as defined in Equation~(\ref{eq:projection}), so $\bar{\mathbf{S}}_{i}(\bm{\phi})\mathbf{M}'\bar{\mathbf{q}}_{1} \geq 0$. From the definition of $\bar{\mathbf{S}}_{i}(\bm{\phi})$, $\bar{\mathbf{S}}_{i}(\bm{\phi})\mathbf{M}'\bar{\mathbf{q}}_{1} \geq 0$ implies that
\begin{align*}
\tilde{\mathbf{S}}_{i}(\bm{\phi}) \mathbf{M}'\bar{\mathbf{q}}_{1} &\geq \tilde{\mathbf{S}}_{i}(\bm{\phi})\tilde{\mathbf{F}}(\bm{\phi})'(\tilde{\mathbf{F}}(\bm{\phi})\tilde{\mathbf{F}}(\bm{\phi})')^{-1}\tilde{\mathbf{F}}(\bm{\phi})\mathbf{M}'\bar{\mathbf{q}}_{1} \\
  \Rightarrow \quad \mathbf{S}_{i}(\bm{\phi})\mathbf{K}\mathbf{M}'\bar{\mathbf{q}}_{1} &\geq 0,
\end{align*}
where the last line follows from $\tilde{\mathbf{S}}_{i}(\bm{\phi}) = (\mathbf{K}^{-1}\mathbf{S}_{i}(\bm{\phi})')'$ and $\tilde{\mathbf{F}}(\bm{\phi})\mathbf{M}'\bar{\mathbf{q}}_{1} = \mathbf{0}_{r\times 1}$.\footnote{Since the last $r$ elements of $\mathbf{M}'\bar{\mathbf{q}}_{1}$ are equal to zero, $\mathbf{M}'\bar{\mathbf{q}}_{1}$ lies in the hyperplane spanned by the first $n-r$ basis vectors. From the construction of the basis, any vector within this hyperplane lies within the null space of $\tilde{\mathbf{F}}(\bm{\phi})$, so $\tilde{\mathbf{F}}(\bm{\phi})\mathbf{M}'\bar{\mathbf{q}} = \mathbf{0}_{r\times 1}$.} $\mathbf{q}_{1} = \mathbf{K}\mathbf{M}'\bar{\mathbf{q}}_{1}$ therefore satisfies $\mathbf{S}(\bm{\phi})\mathbf{q}_{1} \geq \mathbf{0}_{s\times 1}$. Since $\tilde{\mathbf{F}}(\bm{\phi}) = (\mathbf{K}^{-1}\mathbf{F}(\bm{\phi})')'$, $\mathbf{M}'\bar{\mathbf{q}}_{1}$ satisfies $\mathbf{F}(\bm{\phi})\mathbf{KM}\bar{\mathbf{q}}_{1} = \mathbf{0}_{r\times 1}$, so $\mathbf{q}_{1} = \mathbf{K}\mathbf{M}'\bar{\mathbf{q}}_{1}$ additionally satisfies $\mathbf{F}(\bm{\phi})\mathbf{q}_{1} = \mathbf{0}_{r\times 1}$. \hfill$\square$

\medskip

\textbf{Proof of Corollary~\ref{cor:nonempty}:} Assume $\bar{\mathbf{q}}_{1} \in \mathbb{S}^{n-r-1}$ satisfies $\bar{\mathbf{S}}(\bm{\phi})\bar{\mathbf{q}}_{1} \geq \mathbf{0}_{s\times 1}$. Then, from Proposition~\ref{prop:equivalence}(b), $\mathbf{q}_{1} = \mathbf{K}\mathbf{M}'\bar{\mathbf{q}}_{1} \in \mathbb{R}^{n}$ satisfies $\mathbf{F}(\bm{\phi})\mathbf{q}_{1} = \mathbf{0}_{r\times 1}$ and $\mathbf{S}(\bm{\phi})\mathbf{q}_{1} \geq \mathbf{0}_{s\times 1}$. Since $\bar{\mathbf{q}}_{1} \in \mathbb{S}^{n-r-1}$, it has unit norm, which implies that $\mathbf{q}_{1}$ also has unit norm, because multiplication by $\mathbf{M}'$ adds $r$ zeros to $\bar{\mathbf{q}}_{1}$ (leaving the norm unchanged) and $\mathbf{K}$ is orthonormal. $\mathbf{q}_{1}$ therefore lies in $\mathbb{S}^{n-1}$. Since $\mathbf{q}_{1}$ satisfies the identifying restrictions and lies in $\mathbb{S}^{n-1}$, it lies in $\mathcal{Q}_{1}(\bm{\phi}|\mathbf{F},\mathbf{S})$, which must therefore be nonempty.

  Now, assume that $\mathbf{q}_{1} \in \mathbb{S}^{n-1}$ satisfies $\mathbf{F}(\bm{\phi})\mathbf{q}_{1} = \mathbf{0}_{r\times 1}$ and $\mathbf{S}(\bm{\phi})\mathbf{q}_{1} \geq \mathbf{0}_{s\times 1}$. By Proposition~\ref{prop:equivalence}(a), $\bar{\mathbf{q}}_{1} = \mathbf{M}\mathbf{K}^{-1}\mathbf{q}_{1} \in \mathbb{R}^{n-r}$ satisfies $\bar{\mathbf{S}}(\bm{\phi})\bar{\mathbf{q}}_{1} \geq \mathbf{0}_{s\times 1}$. Since $\mathbf{q}_{1}$ has unit norm, so does $\mathbf{K}^{-1}\mathbf{q}_{1}$, since $\mathbf{K}^{-1}$ is orthonormal. The last $r$ elements of $\mathbf{K}^{-1}\mathbf{q}_{1}$ are equal to zero, so $\bar{\mathbf{q}}_{1} = \mathbf{M}\mathbf{K}^{-1}\mathbf{q}_{1}$ also has unit norm and thus lies in $\mathbb{S}^{n-r-1}$. Since $\bar{\mathbf{q}}_{1}$ satisfies $\bar{\mathbf{S}}(\bm{\phi})\bar{\mathbf{q}}_{1} \geq \mathbf{0}_{s\times 1}$ and lies in $\mathbb{S}^{n-r-1}$, it lies in $\bar{\mathcal{Q}}_{1}(\bm{\phi}|\bar{\mathbf{S}})$, which must therefore be nonempty. \hfill$\square$

\end{document}